\documentclass[iop]{emulateapj}
\usepackage{graphicx}
\usepackage{amsmath}
\usepackage{rotating}
\usepackage{threeparttable}

\shorttitle{Variability Selected LLAGN in the 4 Ms \mbox{CDF-S}}

\begin{document}

\title{Variability Selected Low-Luminosity Active Galactic Nuclei in the 4 Ms Chandra Deep Field-South\\
}

\author{
M. Young\altaffilmark{1,2}, W. N. Brandt\altaffilmark{1,2}, Y. Q. Xue\altaffilmark{1,2}, M. Paolillo\altaffilmark{3}, 
D. M. Alexander\altaffilmark{4}, F. E. Bauer\altaffilmark{5,6}, B. D. Lehmer\altaffilmark{7,8}, B. Luo\altaffilmark{9}, O. Shemmer\altaffilmark{10}, 
D. P. Schneider\altaffilmark{1,2}, C. Vignali\altaffilmark{11}} 

\altaffiltext{1}{Department of Astronomy \& Astrophysics, 525 Davey Lab, The Pennsylvania State University, University Park, PA 16802, USA}
\altaffiltext{2}{Institute for Gravitation and the Cosmos, The Pennsylvania State University, University Park, PA 16802, USA}
\altaffiltext{3}{Dipartimento di Scienze Fisiche, Universita` Federico II di Napoli, Via Cinthia, 80126 Napoli, Italy}
\altaffiltext{4}{Department of Physics, University of Durham, South Road, Durham, DH1 3LE, UK}
\altaffiltext{5}{Pontificia Universidad Cat\'{o}lica de Chile, Departamento de Astronom\'{i}a y Astrof\'{i}sica, Casilla 306, Santiago 22, Chile}
\altaffiltext{6}{Space Science Institute, 4750 Walnut Street, Suite 205, Boulder, Colorado 80301}
\altaffiltext{7}{The Johns Hopkins University, Homewood Campus, Baltimore, MD 21218, USA}
\altaffiltext{8}{NASA Goddard Space Flight Centre, Code 662, Greenbelt, MD 20771, USA}
\altaffiltext{9}{Harvard-Smithsonian Center for Astrophysics, 60 Garden St. Cambridge, MA 02138 USA}
\altaffiltext{10}{Department of Physics, University of North Texas, Denton, TX 76203, USA}
\altaffiltext{11}{Universita ́ di Bologna, Via Ranzani 1, Bologna, Italy}

%%%%%%%%%%%%%%%%%%%%%%%%%%%%%%%%%%%%%%%%%%%%%%%%%%%%%%%%%%%%%%%%%%%%%%%%%%%%%%%%%%%%%%%%%%%%%%%%%%%%%%%%%%%%
\begin{abstract}

The 4 Ms \emph{Chandra} Deep Field-South (\mbox{CDF-S}) and other deep \mbox{X-ray} surveys have been highly effective at selecting active galactic nuclei (AGN).  However, cosmologically distant low-luminosity AGN (LLAGN) have remained a challenge to identify due to significant contribution from the host galaxy.  We identify long-term \mbox{X-ray} variability ($\sim$month--years, observed frame) in 20 of 92 \mbox{CDF-S} galaxies spanning redshifts $z \approx 0.08-1.02$ that do not meet other AGN selection criteria.  We show that the observed variability cannot be explained by \mbox{X-ray} binary populations or ultraluminous \mbox{X-ray} sources, so the variability is most likely caused by accretion onto a supermassive black hole.  The variable galaxies are not heavily obscured in general, with a stacked effective power-law photon index of $\Gamma_{\mathrm{stack}} \approx 1.93\pm0.13$, and are therefore likely LLAGN.  The LLAGN tend to lie a factor of $\approx$6--80 below the extrapolated linear variability-luminosity relation measured for luminous AGN.  This may be explained by their lower accretion rates.  Variability-independent black-hole mass and accretion-rate estimates for variable galaxies show that they sample a significantly different black-hole mass-accretion rate space, with masses a factor of 2.4 lower and accretion rates a factor of 22.5 lower than variable luminous AGN at the same redshift.  We find that an empirical model based on a universal broken power-law PSD function, where the break frequency depends on SMBH mass and accretion rate, roughly reproduces the shape, but not the normalization, of the variability-luminosity trends measured for variable galaxies and more luminous AGN.  
\end{abstract}

%Commonly used AGN selection criteria such as those employed by the CDF-S thus miss $\approx$35\% of LLAGN (L$_{0.5-8 \mathrm{keV}}$ $<$ 10$^{42}$). 

\keywords{ galaxies: active --- \mbox{X-ray}s: galaxies}

%%%%%%%%%%%%%%%%%%%%%%%%%%%%%%%%%%%%%%%%%%%%%%%%%%%%%%%%%%%%%%%%%%%%%%%%%%%%%%%%%%%%%%%%%%%%%%%%%%%%%%%%%%%%
\section{Introduction}\label{sec:intro}

Observations show that all nearby galaxies with a massive bulge component host supermassive black holes (SMBHs) \citep[e.g.,][]{FerrareseFord05,Gultekin09}. SMBHs accreting near the Eddington limit ($L$/$L_{Edd}$ $\sim$ 0.1--1) are visible as luminous active galactic nuclei (AGN) that often outshine their host galaxies. Models of AGN lifetime, constrained by observed Eddington ratio distributions, suggest that SMBH growth is dominated by this luminous phase, lasting $\sim$ a few $\times$ 10$^8$ years \citep[e.g.,][]{Marconi04, Shankar04, HopkinsHernquist09}.

Observations including constraints on the sizes of ionized ``bubbles'' around quasars \citep[e.g.,][]{Jakobsen03, Goncalves08} and the lengths of relativistic jets and radio lobes \citep[e.g.,][]{Scheuer95, Blundell99} suggest that the \emph{episodic} lifetime of luminous activity is similar to the total lifetime, implying that a SMBH is triggered to the luminous AGN phase no more than a few times.  SMBHs therefore spend significant amounts of time in quiescent or low-activity phases, which may contribute up to $\sim$20\% of overall SMBH growth \citep{HopkinsHernquist09}.  A SMBH accreting at lower rates ($L$/$L_{Edd}$ $\ll$ 0.1) will appear as a low-luminosity AGN (LLAGN).  LLAGN share several properties with more luminous AGN, including similar \mbox{X-ray} spectral shapes \citep[e.g.,][]{Younes11} and similar radio-loud fractions  and luminosity-dependent, intrinsic \mbox{X-ray} to optical flux ratios \citep[e.g.,][]{Maoz07}. A more complete census of LLAGN is important for understanding SMBH accretion history, but the relative significance of the host galaxy in LLAGN makes a full census of accretion activity a challenge.

Deep \mbox{X-ray} surveys have been effective at selecting a wide variety of AGN, including luminous, unobscured AGN as well as faint and/or obscured AGN \citep[e.g., see][for a review]{BrandtHasinger05}.  \mbox{X-ray} selection criteria usually include cuts on \mbox{X-ray} luminosity and \mbox{X-ray} spectral shape.  Multi-wavelength data further aid \mbox{X-ray} selection by allowing selection via, for example, excess \mbox{X-ray} emission compared to what is expected from optical flux \citep[e.g.,][]{Hornschemeier03} or radio luminosity \citep[e.g.,][]{Alexander05}.

The above methods have been successful in selecting a wide variety of AGN, but nevertheless miss certain populations, such as very heavily obscured AGN and LLAGN \citep[e.g.,][]{Bauer04, Alexander05, Lehmer08}.  While heavily obscured AGN can often be selected in the IR \citep[e.g.,][]{Houck05, Martinez-Sansigre05, Alexander08}, the spectral energy distributions (SEDs) of LLAGN are likely dominated by the host galaxy in other bands.  Even in \mbox{X-ray}s, \mbox{X-ray} binaries, ultraluminous \mbox{X-ray} sources, and hot gas will provide significant contributions to the overall power output. Simulated \emph{Chandra} observations of nearby low-luminosity Seyfert nuclei artificially shifted to $z \sim 0.3$ show that LLAGN would exhibit \mbox{X-ray} luminosities, spectral shapes, and \mbox{X-ray}-to-optical flux ratios consistent with those of normal or optically bright/\mbox{X-ray} faint galaxies \citep{Peterson06}.  By relying on such criteria, deep \mbox{X-ray} surveys may be underestimating AGN fractions. 

Variability potentially provides a useful indicator of whether an extragalactic \mbox{X-ray} source, classified as a galaxy by other means, harbors an AGN.  Variability is a defining characteristic of AGN and has long been used as an AGN selection technique \citep[e.g.,][]{vandenBergh73}.  Numerous studies have used optical variability to select AGN from deep surveys such as the 1 Ms \mbox{CDF-S}, the Subaru/XMM-Newton Deep Field, and the GOODS North and South Fields \citep[e.g.,][respectively]{Trevese08, Morokuma08, Villforth10,Sarajedini11}. Spectroscopic observations of the 1 Ms \mbox{CDF-S} \citep{Boutsia09} found that 17 of 27 optical variability-selected objects were broad-line AGN; 9 (5) AGN would have been missed if selected by color (\mbox{X-ray} selection).  

Similarly, UV variability has been used successfully to identify LLAGN in galaxies with low-ionization nuclear emission-line regions (LINERs).  LINERs have been found in the nuclei of a large fraction of nearby galaxies \citep[e.g.,][]{Ho97, Kauffmann03}, but these regions could be ionized by either massive star clusters or low accretion-rate AGN.  \emph{HST} imaging has found that $\sim$25\% of LINERs are associated with compact ($\lesssim$ few pc) UV sources \citep{Maoz95, Barth98}.  A study of LINERs with compact nuclear UV sources found significant variability in 15 of 17, indicating the presence of an AGN \citep{Maoz05}.
%, with max-to-min amplitudes of a few percent to 50\%, 
%; ``UV-dark" LINERs may also be associated with dust-obscured UV sources (Pogge et al. 2000)

Deep X-ray surveys are able to detect variability in moderate-luminosity/high-redshift AGN \citep[e.g.,][]{Almaini00,Paolillo04,Papadakis08b}.  The 4 Ms \emph{Chandra} Deep Field-South \citep{Xue11}, the deepest \mbox{X-ray} survey to date, allows a preliminary classification of AGN on the basis of several observed quantities (see \S\ref{sec:overview} for details).  This paper utilizes \mbox{X-ray} variability techniques to search for AGN missed by these criteria.  With 4 Ms of exposure time spanning 10.8 years for 466 good-quality sources (see \S\ref{sec:fluxvar}), variability can be detected in sources with time-averaged fluxes as faint as $F_{0.5-8 \mathrm{keV}} \approx 5\times10^{-17}$~ergs~cm$^{-2}$~s$^{-1}$.  We use a cosmology with H$_0$ = 70.4 km s$^{-1}$ Mpc$^{-1}$, $\Omega_M$ = 0.272, and $\Omega_\Lambda$ = 0.728 \citep[e.g.,][]{Komatsu11}.

%%%%%%%%%%%%%%%%%%%%%%%%%%%%%%%%%%%%%%%%%%%%%%%%%%%%%%%%%%%%%%%%%%%%%%%%%%%%%%%%%%%%%%%%%%%%%%%%%%%%%%%%%%%%
\section{Overview of the 4 Ms \mbox{CDF-S} Catalog}\label{sec:overview}

The details of the 4 Ms \mbox{CDF-S} source catalog are available in \citet{Xue11}; we provide a brief summary here. The 4 Ms \mbox{CDF-S}, constructed from 54 \emph{Chandra} observations over 10.8 years, covers an area of 464.5 arcmin$^2$ and reaches highest sensitivities of $F_{0.5-2 \mathrm{keV}}\approx9.1\times10^{-18}$~ergs~cm$^{-2}$~s$^{-1}$ and $F_{2-8 \mathrm{keV}}\approx5.5\times10^{-17}$~ergs~cm$^{-2}$~s$^{-1}$, with multi-wavelength coverage in more than 40 bands from the radio to the UV.  Source candidates are detected using a 10$^{-5}$ false-positive probability threshold in {\sc wavdetect} \citep{Freeman02} and are then pruned using a binomial no-source probability (see Appendix A2 of Weisskopf et al. 2007)\nocite{Weisskopf07} $P$ $<$ 0.004 to obtain a more conservative list of 740 main-catalog sources, all of which are consistent with being point sources. Source extraction and photometry were conducted with {\sc acis extract} (AE; Broos et al. 2010)\nocite{Broos10}.  AE models \emph{Chandra}'s High Resolution Mirror Assembly point spread function (PSF) using the MARX ray-tracing simulator.\footnote{MARX is available at http://space.mit.edu/CXC/MARX/index.html}  The PSF model is used to generate a polygonal extraction region for each source that approximates the $\approx$90\% encircled energy fraction (EEF) contour of a local PSF measured at 1.497~keV.  For $<$6\% of the candidates, the sources are crowded (i.e., the polygonal source regions overlap) and smaller extraction regions that are as large as possible without overlapping ($40-75$\% EEF) are used.  The background is calculated from regions that subtract the contribution from the source of interest and its neighboring sources; the regions are typically a factor $\approx$16 larger than the source-extraction region.  AE merges the individual observations to estimate aperture-corrected, background-subtracted counts and the 1$\sigma$ (asymmmetric) upper and lower statistical errors \citep{Gehrels86}. In this paper, we will use the standard \mbox{X-ray} photometric bands: 0.5--2~keV (soft), 2--8~keV (hard), and 0.5--8~keV (full).  

Though most sources have a relatively small number of counts (median net counts $\approx$ 77), a rough estimate of source spectral shape can be made by relating the band ratio (i.e., the ratio of the count rates in the 2--8~keV and 0.5--2~keV bands) to an effective power-law photon index, $\Gamma_{\mathrm{eff}}$ ($F_\nu$ $\propto$ $\nu^{-\alpha}$ $\equiv$ $\nu^{-\Gamma + 1}$).  For low-count sources where $\Gamma_{\mathrm{eff}}$ cannot be determined reliably, $\Gamma_{\mathrm{eff}}$ is set to 1.4, the stacked (co-added) spectrum of all sources in the \mbox{CDF-S} \citep{Tozzi01,Xue11}, which is consistent with the unresolved spectrum of the cosmic \mbox{X-ray} background \citep{HickoxMarkevitch06}.

Of 740 \mbox{X-ray} sources, 716 (96.8\%) contain matches in at least one of seven optical/near-infrared/radio (ONIR) catalogs: (1) the ESO 2.2 m WFI $R$-band catalog \citep{Giavalisco04}, (2) the GOODS-S \emph{Hubble Space Telescope} ($HST$) version r2.0z $z$-band catalog \citep{Giavalisco04}, (3) the GEMS HST $z$-band catalog \citep{Caldwell08}, (4) the GOODS-S MUSIC catalog \citep{Grazian06}, (5) the MUSYC $K$-band catalog \citep{Taylor09}, (6) the SIMPLE Spitzer/IRAC 3.6$\mu$m catalog \citep{Damen11}, and (7) the VLA 1.4 GHz radio catalog \citep{Miller08}.  

Of 716 \mbox{X-ray} sources with multi-wavelength identifications, 419 (58.5\%) have spectroscopic redshift measurements, collected from \citet{LeFevre04}, \citet{Szokoly04}, \citet{Zheng04}, \citet{Mignoli05}, \citet{Ravikumar07}, \citet{Vanzella08}, \citet{Popesso09}, \citet{Treister09}, \citet{Balestra10}, and \citet{Silverman10}.  A total of 343 (81.9\%) of the 419 spectroscopic redshift measurements are ``secure'', in that they are measured at $\gtrsim$ 95\% confidence levels with multiple secure spectral features.  668 (93.3\%) sources have high-quality, accurate ($|\Delta z |$/($1+z$) $\approx$ 6.5\%) photometric-redshift measurements from at least one of three photometric-redshift catalogs: \citet{Cardamone10}, \citet{Luo10}, and \citet{Rafferty11}.  The positions of primary ONIR counterparts were cross-matched with the photometric-redshift catalogs using a matching radius of 0.5$^{\prime\prime}$, resulting in a false-match probability of $\lesssim$1\%. Subsequent spectroscopic observations published in the Arizona CDF-S Environment Survey (ACES; Cooper et al. 2012)\nocite{Cooper12} catalog show with a blind test that errors on the photometric redshifts are $\lesssim$1\%. 

The 4 Ms \mbox{CDF-S} \mbox{X-ray} sources were classified as AGN by the following criteria: \\
$\indent$$\bullet$ High luminosities: $L_{0.5-8 \mathrm{keV}}$ $\geq3\times10^{42}$~ergs~s$^{-1}$, where the rest-frame luminosity has been corrected for Galactic and intrinsic absorption.\\
$\indent$$\bullet$ Hard spectra: A source with $\Gamma_{\mathrm{eff}}$ $<$ 1.0 is identified as a heavily obscured AGN.\\
$\indent$$\bullet$ High \mbox{X-ray}-to-optical flux ratios: log($F_X$/$F_R$)~$>~-1$, where $F_X$ = $F_{0.5-8 \mathrm{keV}}$, $F_{0.5-2 \mathrm{keV}}$, or $F_{2-8 \mathrm{keV}}$ and $F_R$ is the $R$-band flux. \\
$\indent$$\bullet$ Excess \mbox{X-ray} emission compared to that expected from star formation: $L_{0.5-8 \mathrm{keV}} > 3\times$($8.9\times10^{17}$ $L_R$) \citep{Alexander05}, where $L_R$ is the 1.4 GHz monochromatic luminosity in W Hz$^{-1}$.\\
$\indent$$\bullet$ An indication of broad emission lines in the optical spectrum.
%Note that the criteria requiring multiwavelength detection are only applied when the \mbox{X-ray} source matches the specified ONIR counterpart (e.g., the $L_{0.5-8 \mathrm{keV}} > 3\times$($8.9\times10^{17}$ $L_R$) is only applied to sources detected in the radio). -- This may be important to mention re: Dave's plot, but no need right now.

Stars were classified by cross-matching \mbox{X-ray} sources (using the ONIR counterpart positions) with (1) the spectroscopically identified stars in
\citet{Szokoly04}, \citet{Mignoli05}, and \citet{Silverman10}; (2) the likely stars with stellarity indices greater than 0.7 in the GEMS \emph{HST} catalog \citep{Caldwell08}; and (3) the likely stars with best-fit stellar templates in the MUSYC photometric-redshift catalog \citep{Cardamone10}, using a matching radius of 0.5$^{\prime\prime}$.  

\mbox{X-ray} sources not identified as an AGN or a star were classified as galaxies.

Rest-frame 0.5--8~keV luminosities are calculated for all sources.  For AGN, which make up the vast majority of \mbox{CDF-S} sources, the luminosity is corrected for Galactic and intrinsic absorption.  AGN not detected in the full band have an upper limit on the \mbox{X-ray} luminosity based on the 3$\sigma$ Poisson error on the counts.  For galaxies, the intrinsic-absorption correction and $K$-corrections may not be appropriate.  Of 92 \mbox{CDF-S} galaxies with ``good quality'' observations (see \S\ref{sec:fluxvar}), 78 are not detected in the hard band and have poorly determined photon indices.  All galaxies are detected in the soft-band.  We perform a stacking analysis for galaxies with $<$ 150 net counts (87 of 92) following the procedure described in \S\ref{sec:stack}.  The resulting average $\Gamma_{\mathrm{stack}}$ = 1.90$\pm$0.08 gives $L_{0.5-8 \mathrm{keV}}$/$L_{0.5-2 \mathrm{keV}}$ = 2.35.  We calculate the rest-frame 0.5--8~keV luminosity for \mbox{CDF-S} galaxies as $L_{0.5-8 \mathrm{keV}} = 4 \pi d_L^2 \times 2.35 \times f_{0.5-2 \mathrm{keV}} (1+z)^{\Gamma_{\mathrm{stack}} - 2}$.  Since galaxy \mbox{X-ray} emission is typically unabsorbed, we do not apply any correction for intrinsic absorption.

%%%%%%%%%%%%%%%%%%%%%%%%%%%%%%%%%%%%%%%%%%%%%%%%%%%%%%%%%%%%%%%%%%%%%%%%%%%%%%%%%%%%%%%%%%%%%%%%%%%%%%%%%%%%
\section{Testing for \mbox{X-ray} Variability}\label{sec:fluxvar}

We perform two quality cuts before conducting variability analysis.  First, we exclude the catalog sources with off-axis angles greater than 8$^\prime$ to ensure that sources will have sufficient coverage ($>$50 of 54 observations) throughout the various \mbox{CDF-S} pointings. To ensure accurate variability measurements, we also require that each source has at least 20 net counts in the 0.5--8.0 keV band (i.e., at least 5 counts on average per epoch, as defined below).  These quality cuts result in a total sample of 466 \mbox{CDF-S} sources: 369 classified as AGN, 92 classified as galaxies, and 5 classified as stars in the \citet{Xue11} catalog.  All of these AGN and galaxies have measured spectroscopic or photometric redshifts.  The 92 sources classified as galaxies may nevertheless contain an AGN, as would be indicated by \mbox{X-ray} variability, and make up the sample investigated in this paper.

We divide the \mbox{CDF-S} observations into four epochs, each containing $\sim$1~Ms of integration: 2000 (943.1 ks), 2007 (967.7 ks), 2010a (March--May; 1015.5 ks), and 2010b (May--July; 944.9 ks).  As in \citet{Xue11}, we merged observations within each epoch and, for a given source position from the \mbox{CDF-S} catalog, measured the source and background counts and flux over three observed-frame energy bands: 0.5--8 keV, 0.5--2 keV, and 2--8 keV.  A source is considered variable if the variability observed between observations is greater than that expected from Poisson statistics, with a probability threshold of 5\% that the variability is due to noise alone.  (The choice of probability threshold is discussed further below.)  To check whether a source can be considered variable, we calculate the quantity:\\
\begin{equation} X^2 = \sum\limits_{i=1}^N \frac{(x_i - \mu)^2}{\sigma_i^2}, \end{equation}
where $N=4$ is the number of epochs, $x_i$ is the photon flux (background-subtracted counts with units of cm$^{-2}$~s$^{-1}$) in a given epoch, $\mu$ is the mean photon flux over all epochs, and $\sigma_i^2$ is the error squared on the photon flux for the $i^{th}$ epoch.  The photon flux is calculated by dividing the full-band (0.5--8.0 keV) net counts by the exposure time and the mean effective area across the source aperture.  The \citet{Gehrels86} approximation gives the error on the net counts, which is propagated to obtain the error on the photon flux. Since this error is significantly asymmetric for low-count ($\lesssim 15$ counts) sources, we average the upward and downward error bars for these objects to obtain the average error $\sigma_i$.  (The same method is applied in the Monte Carlo simulations below.) 
% "asymmetric" = hi/lo ratio < 30%

For large photon fluxes, the $X^2$ statistic follows a $\chi^2$ distribution, and any source with $X^2$ $>$ 7.82 (for 3 degrees of freedom) has a probability $P_{X^2}$ $<$ 0.05 (i.e., 95\% confidence level) that the variability is due to random noise.  However, at low count rates, the error on the photon flux is not Gaussian.  Since errors in the low-count regime are larger than expected from a Gaussian distribution, the resulting $X^2$ statistic is smaller than expected and does not follow the $\chi^2$ distribution (see Fig. 2 of \citealt{Paolillo04}).  

We therefore constructed a Monte Carlo simulation to determine the distribution the $X^2$ statistic should follow for each source, similar to the procedure followed by \citet{Paolillo04}.  We first scaled the total observed source and background counts for each source, obtained from the full 4 Ms observation, to the exposure time and effective area for a given epoch.  This procedure generates the source and background counts expected in each epoch if the source and background were constant over time, and it accounts for fluctuations in the background that will affect low-count sources.  To simulate the variance expected from noise, Poisson distributions were defined using the expected source and background counts as the mean values.  We then simulated 1,000 observations of each source by repeatedly drawing the expected counts from the Poisson distributions. For each simulated observation, we calculated the photon flux for four epochs and calculated $X^2$ as defined above. Asymmetric errors on the source and background counts are obtained from \citet{Gehrels86} and are propagated to get the error on the photon flux.  The observed $X^2$ can then be compared to the simulated distribution to determine the probability $P_{X^2}$ that the observed variability is due to Poisson noise.  Spurious sources of variability are negligible, since effective exposure maps are calculated separately for each observation, taking into account issues such as vignetting, CCD gaps, bad pixels, bad columns, and \emph{Chandra}'s spatial- and time-dependent quantum efficiency degradation.

While using a more conservative $P_{\mathrm{crit}}$ = 1\% on our dataset would result in fewer false positives, it would also eliminate a similar number of truly varying sources.  For example, in the sample of 92 galaxies, $P_{\mathrm{crit}}$ = 5\% results in 20 variable sources (see \S\ref{sec:var}), of which 4.6 are expected to be false positives. Reducing $P_{\mathrm{crit}}$ to 1\% results in 13 variable sources, of which 0.9 is expected to be a false positive.  So while the more conservative critical value eliminates $\approx$4 false positives, it also eliminates $\approx3-4$ truly varying sources.  The $P_{X^2}$ values listed in Table 2 can be used to screen the sources further as desired.

%LOOK AT VARIABILITY IN DIFFERENT BANDS?  WILL NEED TO ACCOUNT FOR DIFFERENCE IN COUNTS IF WANTING TO COMPARE...

%\subsection{Kolmogorov-Smirnov Test of Intra-Observation Variability}
%(USE K-S TEST TO TEST FOR FALSE POSITIVES? E-MAIL PAT BROOS.)

%%%%%%%%%%%%%%%%%%%%%%%%%%%%%%%%%%%%%%%%%%%%%%%%%%%%%%%%%%%%%%%%%%%%%%%%%%%%%%%%%%%%%%%%%%%%%%%%%%%%%%%%%%%%
\section{Galaxies with AGN-like Variability}\label{sec:var}
%Variability-selection vs. Other Selection Methods

\begin{figure}[b]
\centering
\includegraphics[width=5.5in]{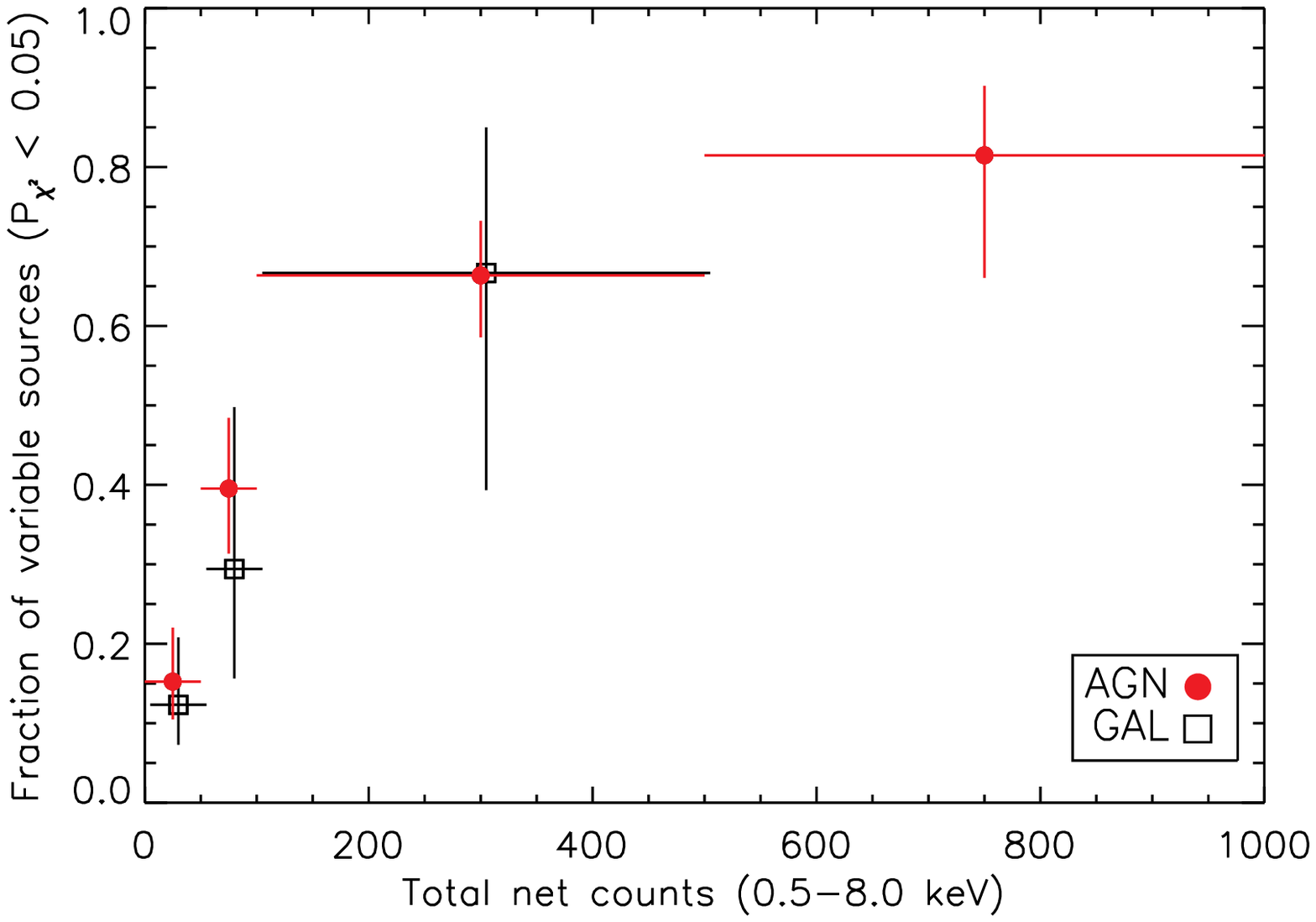}
\caption{The fraction of variable AGN (red circles) and galaxies (black squares) vs. net counts, where the background-subtracted (net) counts are measured in the 0.5--8.0~keV band over 4 Ms. The binomial errors on the fraction at the 90\% confidence level are calculated from Cameron (2011).  The error bar on the net counts represents the bin size.  The points are slightly offset for clarity.}
\label{fig:fracvar}
\end{figure}

Of the 369 \mbox{CDF-S} sources classified as AGN that meet both the off-axis angle ($\theta$ $<$ 8$^\prime$) and count (total net counts $>$ 20) requirements, 50.1\% exhibit significant flux variability ($P_{X^2}$ $<$ 0.05) on $\sim$month--year timescales.  For the 178 AGN with more than 100 counts, 74.2\% exhibit significant variability. The basic diagnostic plot in Figure~\ref{fig:fracvar} shows the rise in the AGN variable fraction with total net counts.  The plot demonstrates that, given sufficient counts to detect it, variability is a near-ubiquitous property of faint, \mbox{X-ray} selected AGN, even in the case of significant obscuration: $\approx$70\% of the \mbox{CDF-S} sample consists of obscured AGN \citep{Xue11}, and 47.5\% (51.5\%) of obscured (unobscured) AGN are significantly variable.  The variable fractions are consistent with the results from the 1 Ms \mbox{CDF-S} \citep{Paolillo04}, although obtained with different temporal sampling and down to much fainter fluxes. 

\begin{figure}[btp!]
\centering
\includegraphics[width=5.5in]{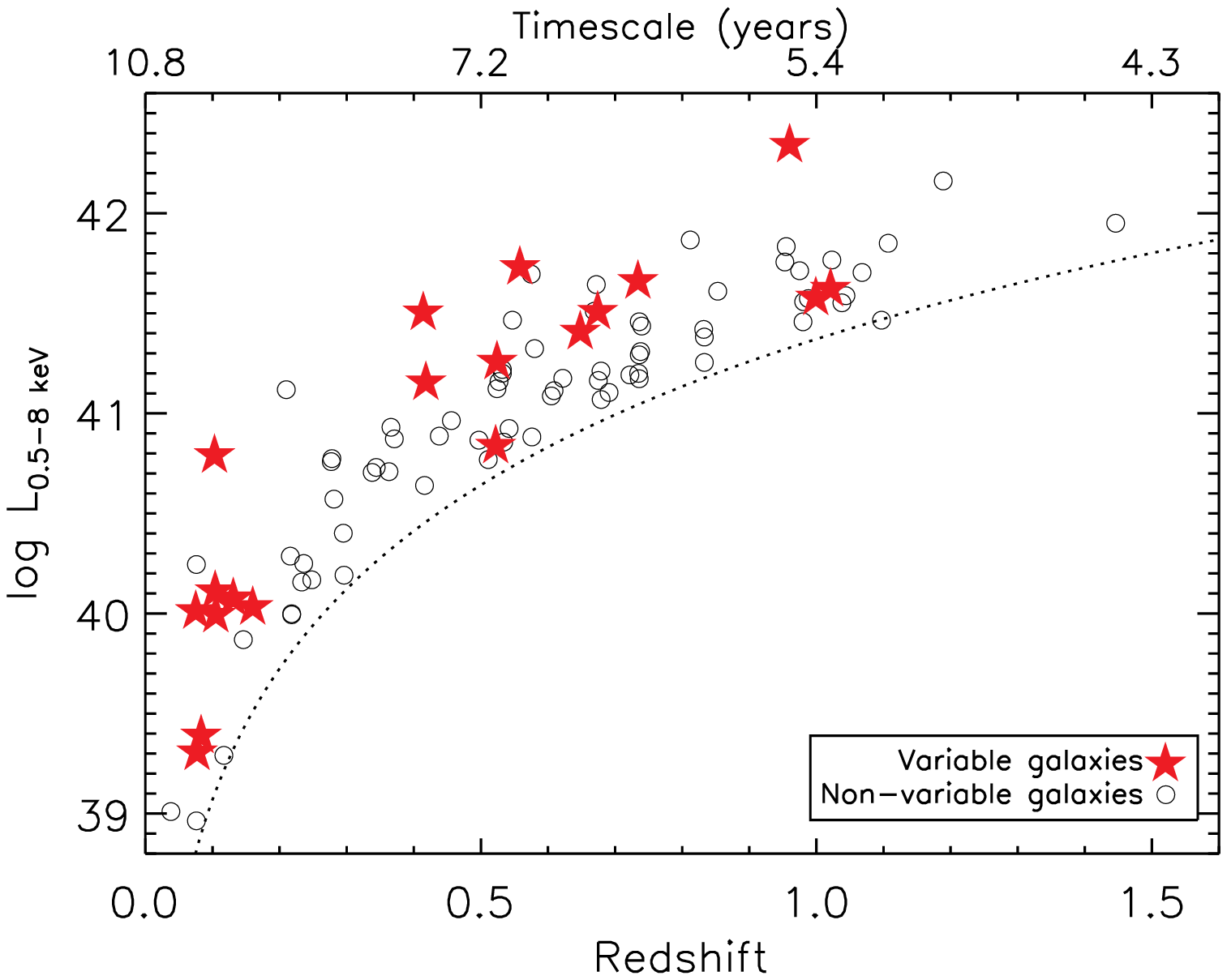}
\caption{The 0.5--8~keV luminosity vs. redshift distribution for \mbox{CDF-S} galaxies meeting the quality criteria of \S\ref{sec:fluxvar} (open circles).  Galaxies with significant variability are marked as red stars.  The \mbox{CDF-S} on-axis flux limit for 20 net counts ($F_{0.5-8 \mathrm{keV}} \approx 4.7\times10^{-17}$~ergs~cm$^{-2}$~s$^{-1}$) is plotted as a dotted line.  The upper $x$-axis shows the maximum rest-frame timescale in years, where the maximum observed-frame timescale of the \mbox{CDF-S} is 10.8 years.    }
\label{fig:Lz}
\end{figure}

\begin{figure*}[btp!]
\centering
\includegraphics[width=3in]{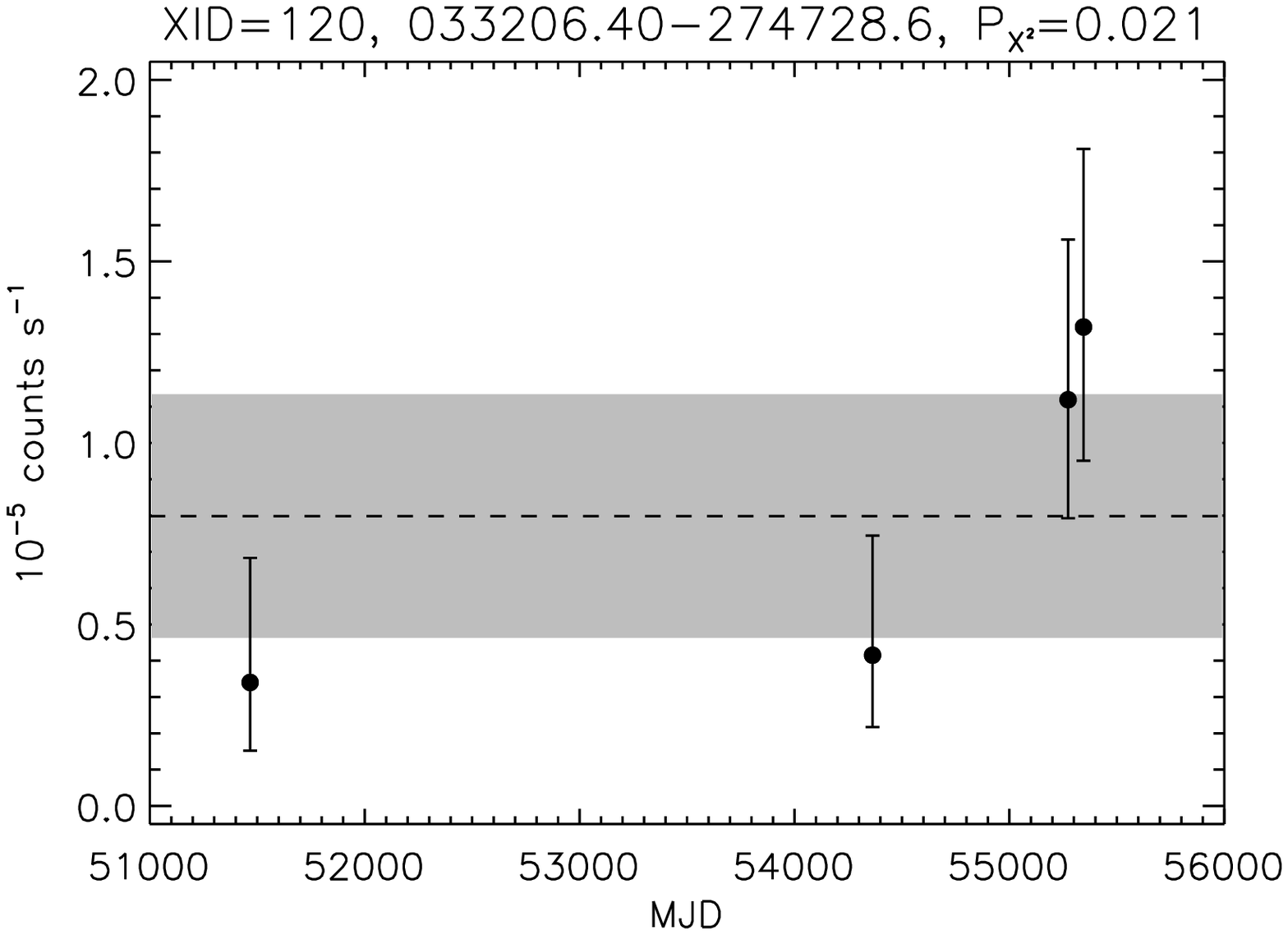}  %File definitions in LC/convert_plots.txt
\includegraphics[width=3in]{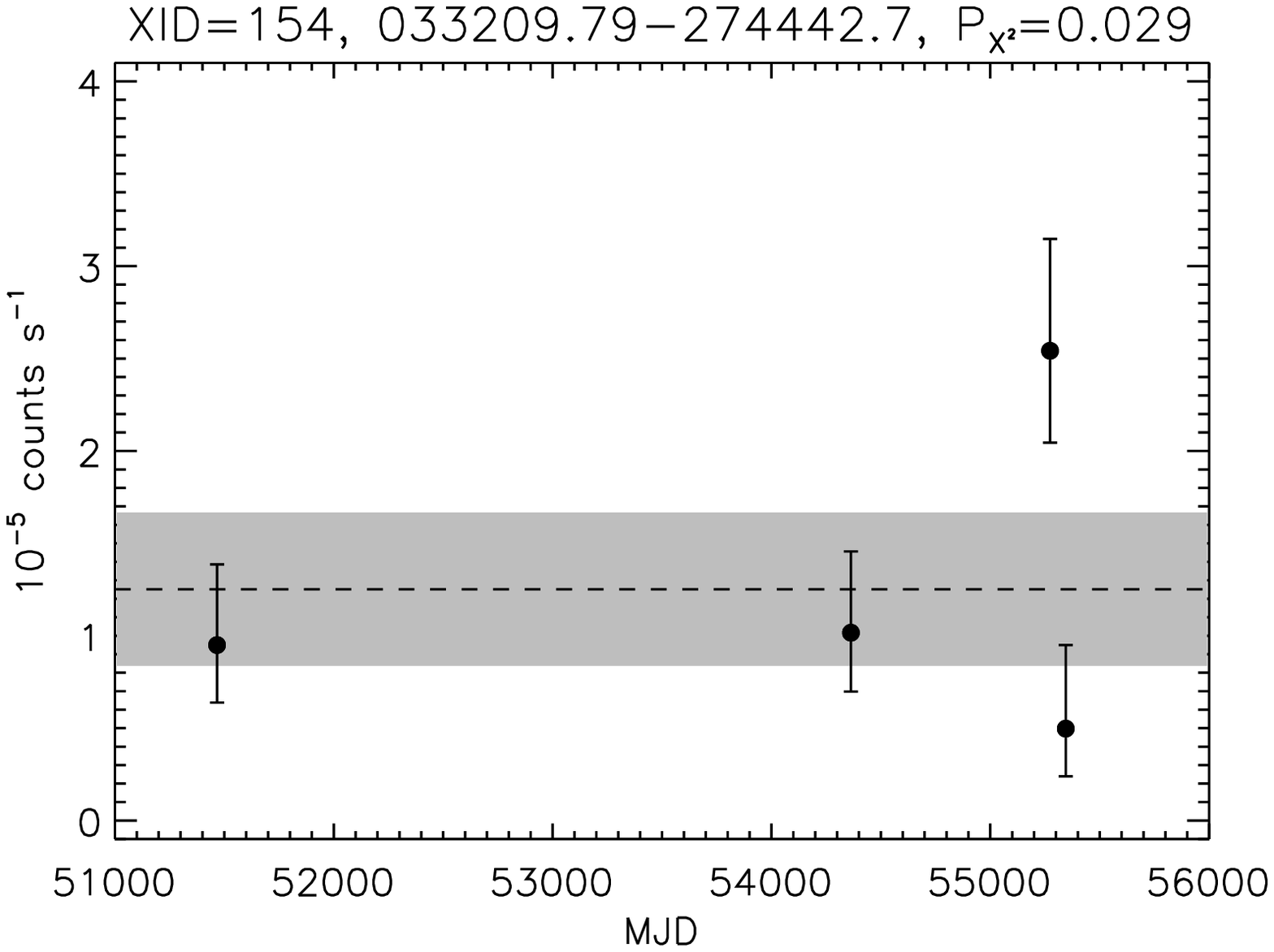}
\includegraphics[width=3in]{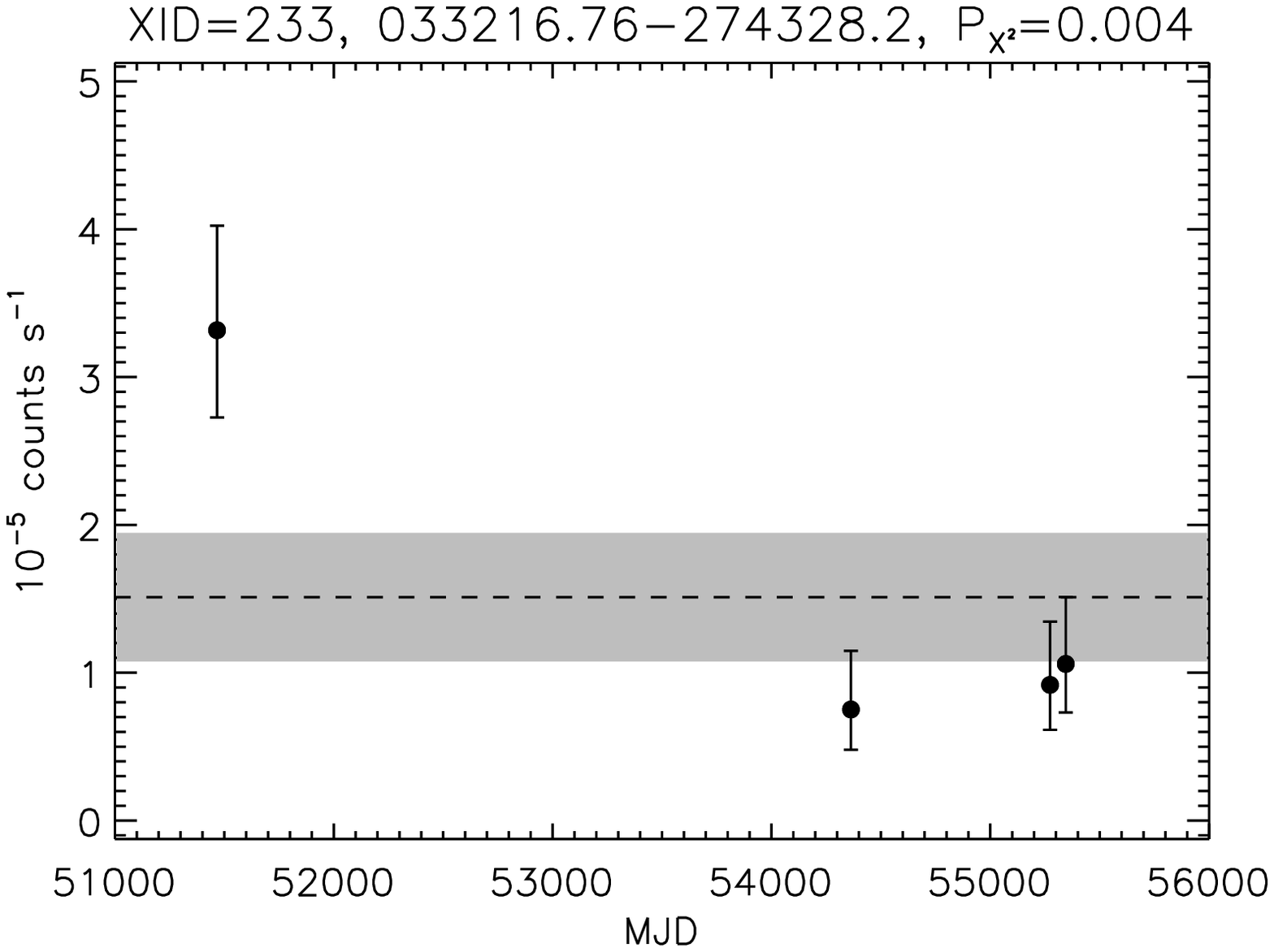}
\includegraphics[width=3in]{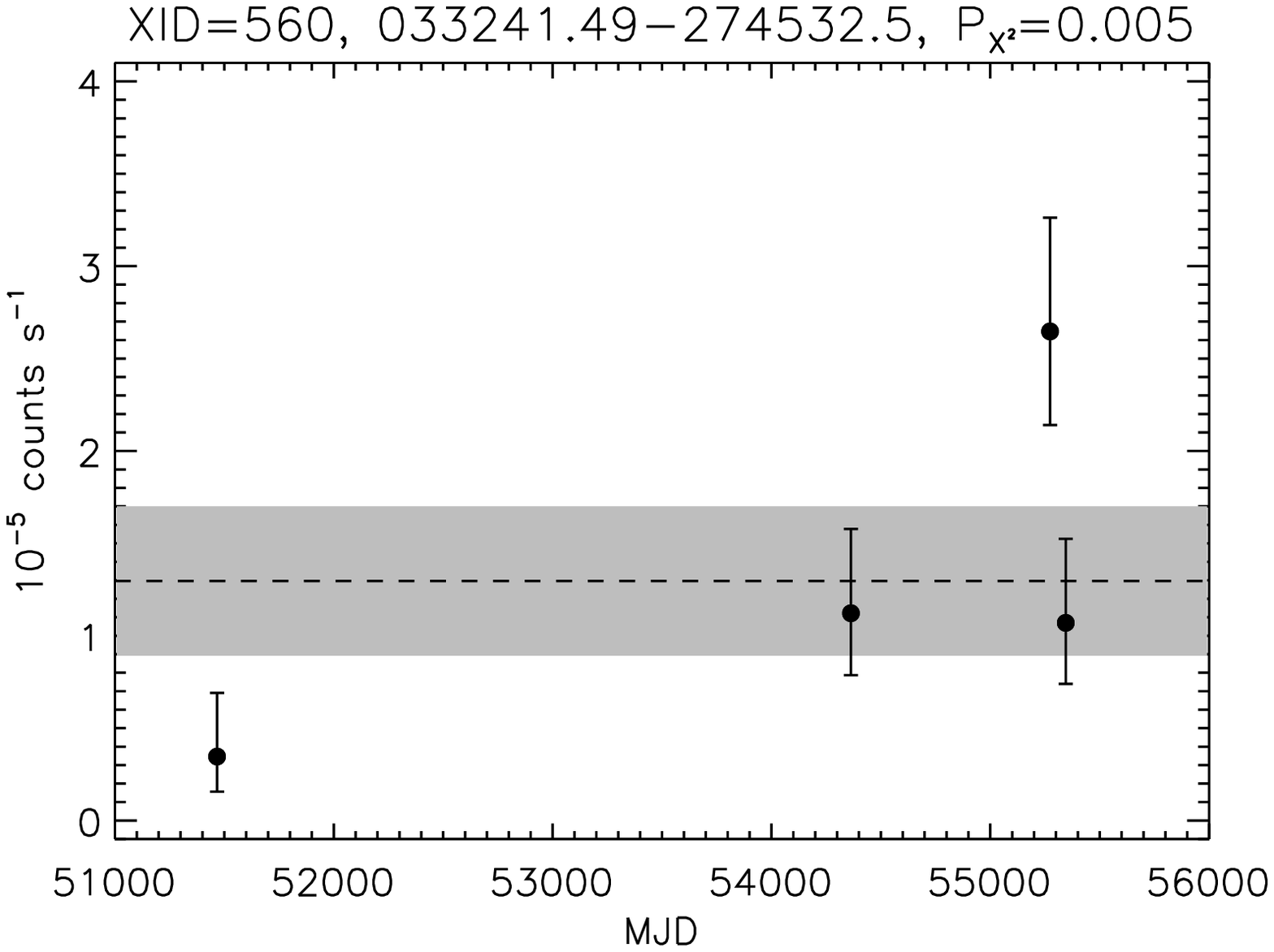}
\includegraphics[width=3in]{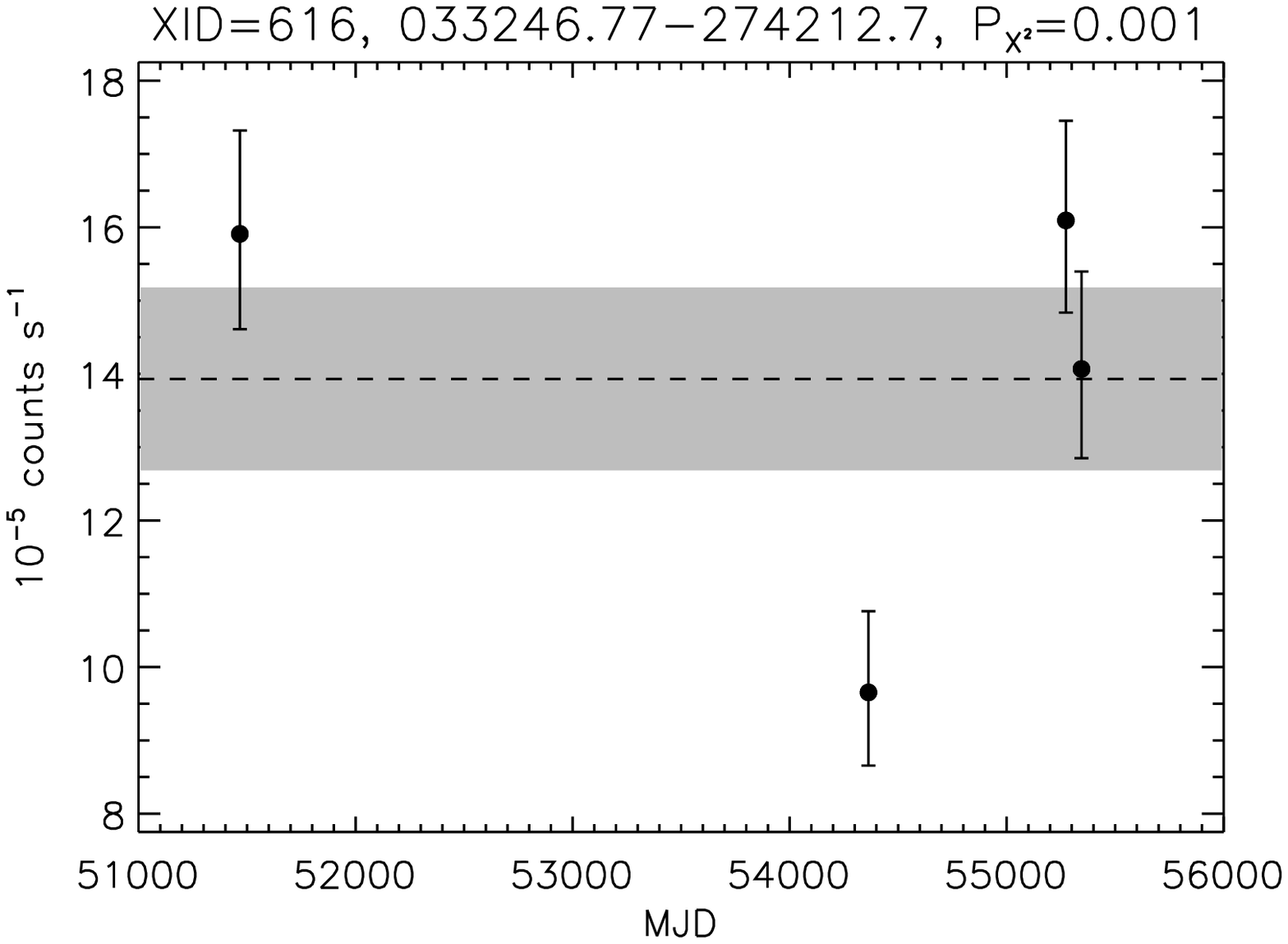}
\includegraphics[width=3in]{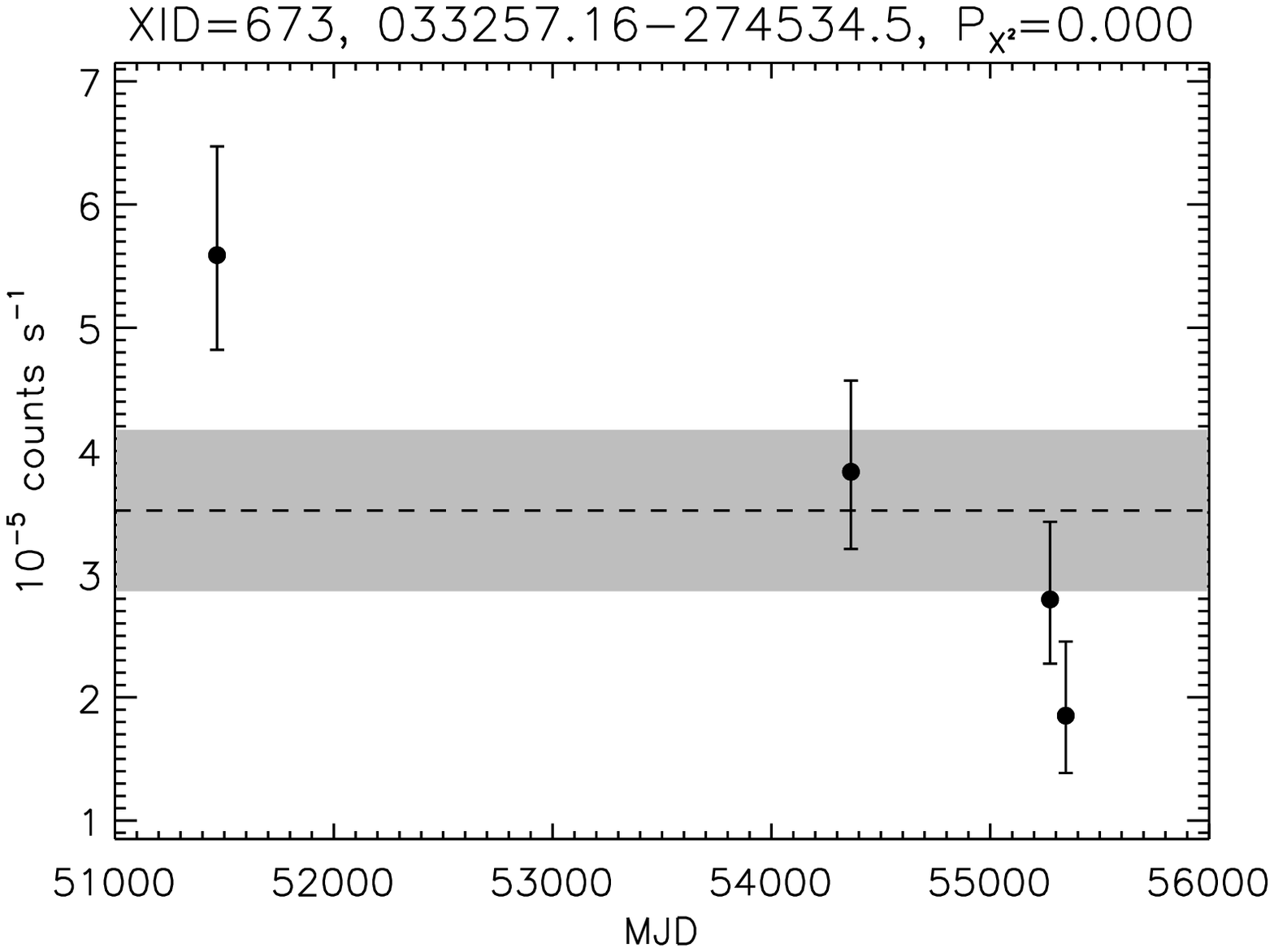}
\caption{Light curves for 6 representative \mbox{CDF-S} sources classified as galaxies that exhibit significant variability. Full-band (0.5--8 keV, observed-frame) counts have been grouped into the four epochs and asymmetric errors on the count rates were calculated via \citet{Gehrels86}.  The mean count rate is overplotted as a dashed line, and the error on the mean is shown as the grey shaded area. Each plot lists the $P_{X^2}$ values and source names.}
\label{fig:lc}
\end{figure*}

In sources classified as galaxies, the variable fraction is significant at low counts and equals that of AGN at higher counts (Figure~\ref{fig:fracvar}). Table 1 describes the columns of Table 2, which lists the attributes of the 20 variable and 72 non-variable galaxies.  (Variability properties of CDF-S AGN will be covered in the forthcoming Paolillo et al., in preparation.)  The luminosity-redshift distribution of variable and non-variable galaxies is shown in Figure~\ref{fig:Lz}; almost all lie below $z \sim 1$. Spectroscopic redshifts are available for 18 of 20 variable galaxies and for 61 of 72 non-variable galaxies.  Photometric redshifts are available for the remaining sources. Six example light curves (background-subtracted count rates in the observed-frame 0.5-8 keV band vs. MJD) representative of the sample as a whole are shown in Figure~\ref{fig:lc}. 

\begin{figure}[btp!]
\centering
\includegraphics[width=5.5in]{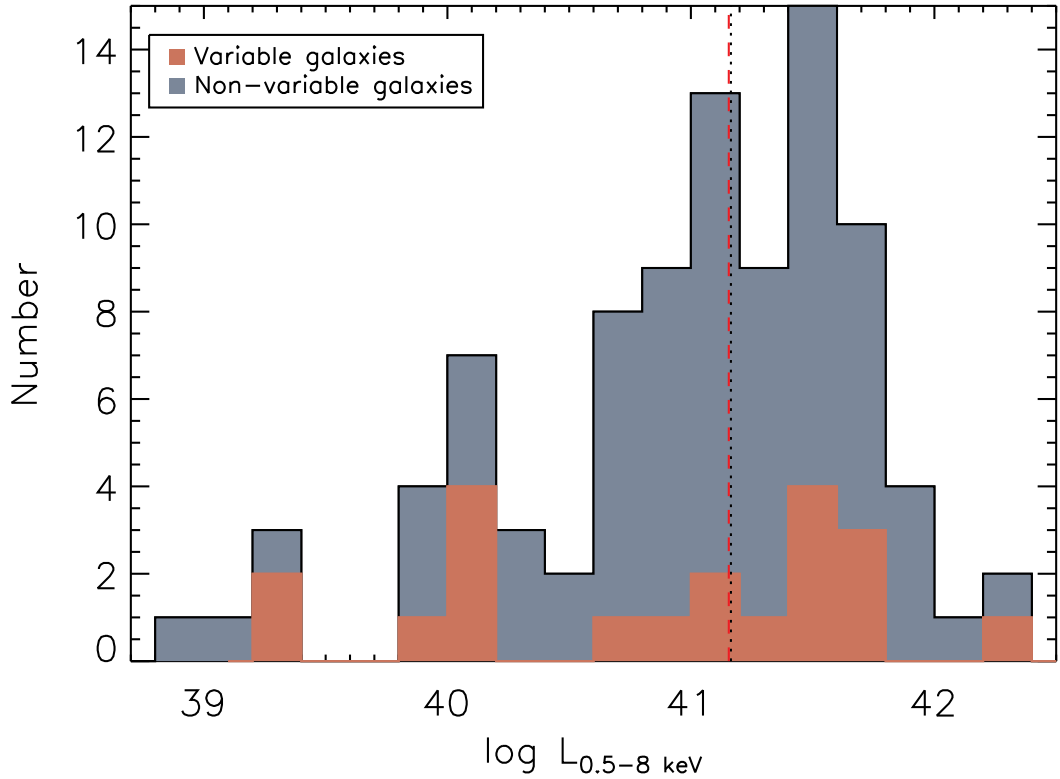}
\caption{The distribution of \mbox{X-ray} luminosity ($L_{0.5-8 \mathrm{keV}}$) for all \mbox{CDF-S} sources classified as galaxies (dark grey histogram) and for those exhibiting significant variability (red histogram). The vertical lines show the median values for all galaxies (black, dotted) and for variable galaxies (red, dashed). A K-S test shows no significant difference between the distributions.}
\label{fig:Lx_gal_histo}
\end{figure}

The \mbox{X-ray} luminosity distributions of all \mbox{CDF-S} galaxies and those exhibiting significant variability are shown in Figure~\ref{fig:Lx_gal_histo}. A K-S test shows that the two samples are consistent with being drawn from the same parent population ($P_{KS}$ = 57\%), and the variable fraction does not show a significant dependence on \mbox{X-ray} luminosity below $L_{0.5-8 \mathrm{keV}}$ $\sim$ 10$^{43}$~ergs~s$^{-1}$.  The possibility that AGN-related variability may go undetected in galaxies is discussed in \S\ref{sec:nxsvar}.

\begin{figure}[btp!]
\centering
\includegraphics[width=5.5in]{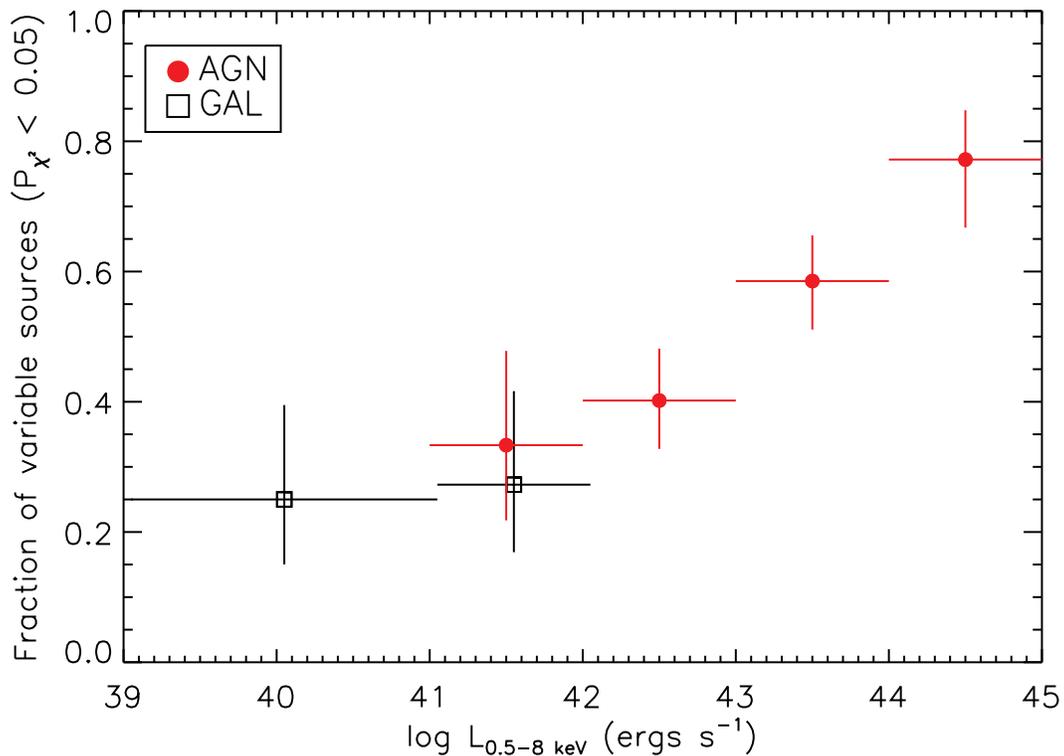}
\caption{The fraction of variable AGN (red circles) and galaxies (black squares) vs. \mbox{X-ray} luminosity. Error bars show the binomial errors at 90\% significance level.  The error bar on luminosity represents the bin size.  The points are slightly offset for clarity.}
\label{fig:fracvar_Lx}
\end{figure}

We briefly compare AGN selection based on variability to the following selection methods: (1) \mbox{X-ray} luminosity cuts, (2) the \mbox{X-ray}-to-optical flux ratio, and (3) excess \mbox{X-ray} emission compared to that expected from star formation, based on the radio luminosity.  Figure~\ref{fig:fracvar_Lx} shows the fraction of variable sources vs. \mbox{X-ray} luminosity.  Below $L_{0.5-8 \mathrm{keV}}$ = 10$^{42}$~ergs~s$^{-1}$, a luminosity cut often used for AGN selection in \mbox{X-ray} surveys, the variable fraction remains significant at $20-30\%$.  Of 64 \mbox{CDF-S} galaxies with $L_{0.5-8 \mathrm{keV}}$ $<$ 10$^{42}$~ergs~s$^{-1}$, 17 ($\approx$27\%) are variable.  

\begin{figure}[btp!]
\centering
\includegraphics[width=5.5in]{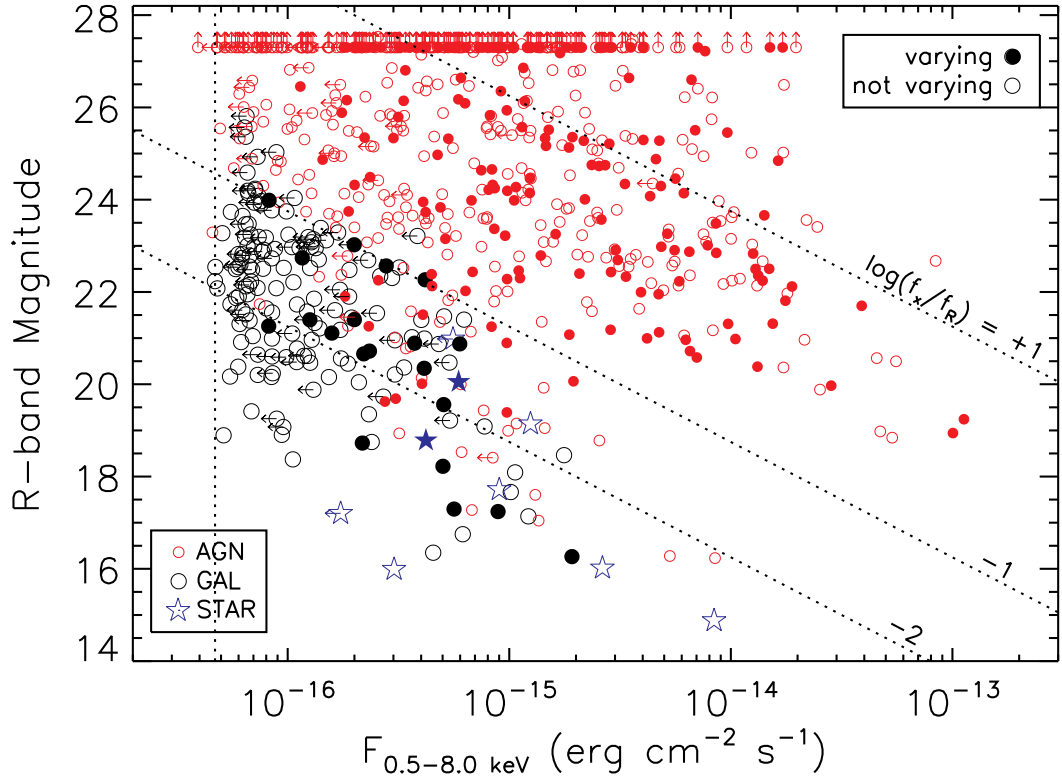}
\caption{$R$-band magnitude vs. full-band \mbox{X-ray} flux for sources in the main \mbox{CDF-S} catalog of \citet{Xue11}, where filled symbols mark sources with significant variability ($P_{X^2}$ $<$ 0.05). Sources classified as AGN, galaxies, and stars are marked with red, black, and blue symbols, respectively. Constant flux ratios are marked with diagonal lines: from top to bottom, log($F_x$/$F_R$) = $+1$, $-1$, and $-2$. The vertical dotted line marks the on-axis full-band flux limit for 20 net counts.  The sources with $R$-band magnitude lower limits at the top of the plot ($\approx$25\% of \mbox{CDF-S} sources) have no $R$-band counterpart.}
\label{fig:xue}
\end{figure}

AGN selection via the \mbox{X-ray}-to-optical flux ratio is demonstrated by the $R$-band magnitude vs. \mbox{X-ray} flux plane in Figure~\ref{fig:xue} (cf. Figure 16 in Xue et al. 2011)\nocite{Xue11}. Sources classified as AGN, galaxies, and stars in \citet{Xue11} are marked as small red circles, larger black circles and blue stars, respectively.  Variable sources are marked with filled symbols. The optically bright, \mbox{X-ray} faint region, typically thought to exclude AGN (OBXF; $F_X$/$F_R <-2$; Hornschemeier et al. 2003)\nocite{Hornschemeier03} contains 27 galaxies.   Of these, 6 (22\%) are variable, with \mbox{X-ray} luminosities spanning log $L_{0.5-8 \mathrm{keV}}$ $\approx$ 39.7--41.4. 

AGN may also be selected based on a comparison between \mbox{X-ray} and radio luminosities \citep{Xue11}.  The radio luminosity can be used to predict the \mbox{X-ray} luminosity in star-forming galaxies \citep[e.g.,][and references therein]{Alexander05}, so sources with excess \mbox{X-ray} emission may be classified as AGN.  Of the 17 \mbox{CDF-S} galaxies with radio detections (none of which have excess X-ray emission), three (18\%) are variable.  These objects may have excess radio emission due to strong radio cores rather than star formation.
%Figure \ref{fig:Alex} shows variable sources in the \mbox{X-ray} luminosity vs. radio luminosity plane (cf. Figure 2$b$ in Alexander et al. 2005).  The dotted line shows an estimate of the radio contribution expected from high-mass \mbox{X-ray} binaries; sources above this line are classified as AGN.  

%%%%%%%%%%%%%%%%%%%%%%%%%%%%%%%%%%%%%%%%%%%%%%%%%%%%%%%%%%%%%%%%%%%%%%%%%%%%%%%%%%%%%%%%%%%%%%%%%%%%%%%%%%%%
\subsection{Measuring Variability Strength}\label{sec:nxsvar}

%If wanting max-to-min amplitude instead of max-to-min factor: 0.4--2.1 (median=4.1)
Due to the generally limited photon statistics of the \mbox{CDF-S} galaxy sample, most variable sources must be strongly variable to be detected.  Significantly variable galaxies show maximum-to-minimum flux ratios $R_{\mathrm{max/min}}$ $\approx$ 1.5--9.3 (median = 4.1) over the observed 10.8-year time frame. The smallest max-to-min ratio (1.5) was measured for 033246.77$-$274212.7 (XID = 616), a source with $\gtrsim$500 counts. For most galaxies, total net counts are too low ($\lesssim 100$) to detect variability below a factor of 2--3.  

To address whether the variable galaxy max-to-min ratios sample the average AGN population or only the highly variable ``tip of the iceberg,'' we ran a Monte Carlo simulation. The procedure assumed that the entire galaxy population is significantly variable and simulated the variability expected over 10.8 years of observation.  Following the procedure in \citet{Vaughan03}, we used the \citet{TimmerKoenig95} algorithm to simulate 5,000 light curves based on the mean and variance of the flux for each of the 92 \mbox{CDF-S} galaxies.  For non-variable galaxies, the measured variance represents noise in the measured light curve, which gives an upper limit to the variability that could be present.  The algorithm produces a random, continuously sampled light curve from a given power spectral density (PSD) function, which we assumed to be described by a broken power-law, where the break frequency depends on SMBH mass and accretion rate \citep{McHardy06}.  Since the break frequency lies outside the range of timescales sampled for most variable galaxies (see \S\ref{sec:model}), we simplify the model to a power-law ($P$($f$) $\propto$ $f^{\beta}$) with index $\beta = -1$, as is typical for the low-frequency (long-timescale) PSDs for nearby Seyferts \citep[e.g.,][]{Vaughan03}.  The time units of the light curves are determined by the minimum and maximum timescales input into the simulation; the light curve durations are adjusted according to each source's redshift.  

We resampled/rebinned the light curve using the \mbox{CDF-S} observing pattern and add Poisson noise to the simulated light curve to account for measurement error.  The full simulated light curves was made five times longer than the sampled region in order to produce variation on timescales much longer than those sampled by the data.  This reproduces the effect that very long-timescale (low-frequency) variations have on variability measured over shorter timescales (i.e., ``red noise leak'').

\begin{figure}[b]
\centering
\includegraphics[width=5.5in]{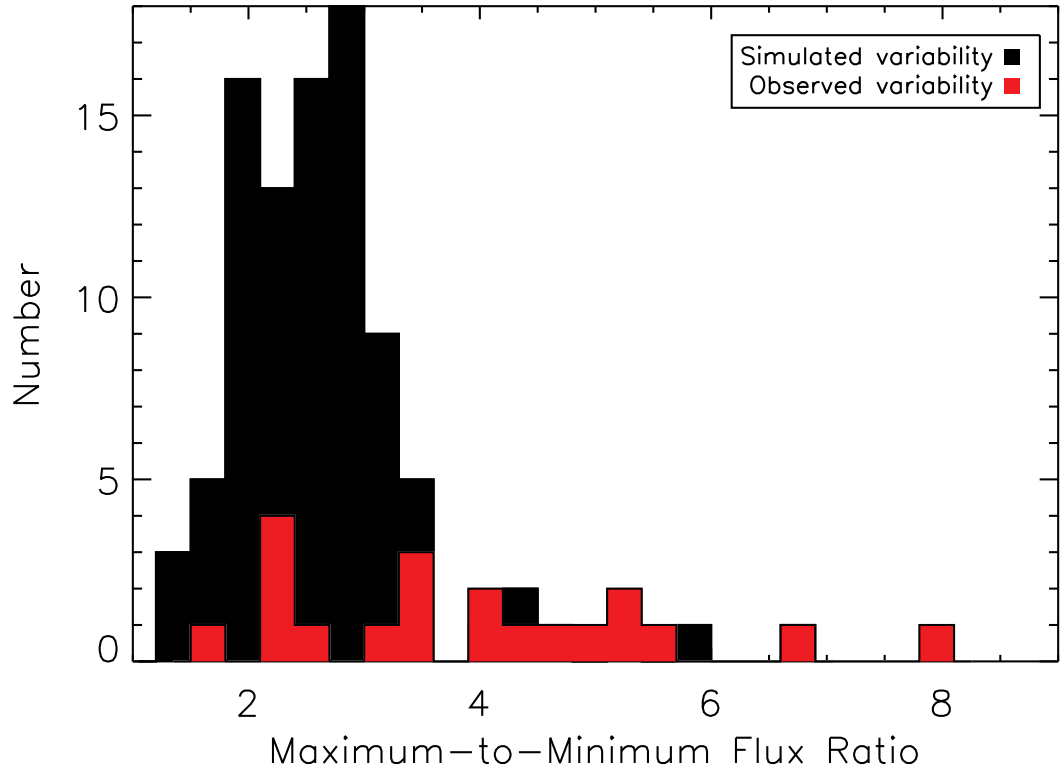}
\caption{The distribution of max-to-min ratios for simulated light curves based on \mbox{CDF-S} galaxies (black histogram) and for measured light curves of variable galaxies (red histogram).}
\label{fig:peaktopeak}
\end{figure}

We compare the simulated distribution of median max-to-min ratios, where the median is calculated over 5,000 trials for each galaxy, to the observed distribution for variable galaxies in Figure~\ref{fig:peaktopeak}.  A K-S test shows that these two populations have a 0.2\% chance of being drawn from the same parent population.  Note that the simulated distribution (black histogram) illustrates an \emph{upper limit} to the variability that could be present in the \mbox{CDF-S}, resulting in a \emph{lower limit} on the detectable fraction of sources.  The \mbox{CDF-S} detects at least $\approx$18\% of sources with max-to-min ratios $>$ 2 and at least $\approx$57\% of sources with max-to-min ratios $>$ 4.  A significant fraction of non-variable galaxies may still host an AGN, but the variability may remain undetected due to low counts.
%The median simulated max-to-min ratio for all 92 galaxies is 2.7, compared to a median of 4.1 for 20 variable galaxies.  

For the galaxies exhibiting significant variability, we calculated the normalized excess variance \citep[e.g.,][]{Vaughan03}, which measures how strongly a source varies in excess of the measurement error.  The excess variance is the integral of a source's PSD function over a given frequency range, which is defined by the light curve's duration (10.8 years, observed-frame) and minimum bin size (4.0 Ms,\footnote{While each \mbox{CDF-S} epoch totals $\sim$1 Ms in integration time, the timescale sampled is significantly longer due to the spread in \emph{Chandra} observations, of which the shortest is 4.0 Ms for the 2007 epoch \citep{Luo08}.} observed-frame).\\
\begin{equation} \sigma^2_{\mathrm{nxs}} = \frac{1}{(N-1)\mu^2} ~ \sum_{i=1}^N (x_i - \mu)^2 - \frac{1}{N \mu^2} ~ \sum_{i=1}^N \sigma_{err,i}^2, \end{equation}
where $\sigma_{err,i}$ is the average of the asymmetric upward and downward measurement errors.  Using the upward (downward) error would overestimate (underestimate) the errors.  Zero excess variance ($\sigma_{\mathrm{nxs}}^2 = 0$) would indicate that the observed count fluctuation is entirely consistent with noise rather than due to intrinsic source variability; due to statistical fluctuations, the excess variance may also be negative in this case.  The $\sigma_{\mathrm{nxs}}^2$ values are listed in Table~2.  Note that the variability amplitude is calculated for \emph{observed-frame} energy bands.  The variable galaxies cover a redshift range $z = 0-1$, so the excess variance will be measured over $0.5-8$~keV at $z$ = 0 to $1-16$ keV at $z$ = 1.  Variability strength is known to depend on energy in some nearby Seyfert galaxies (e.g., Ark 120, MCG--6-30-15, and I Zw 1; Vaughan et al. 2004, Vaughan \& Fabian 2004, and Gallo et al. 2007, respectively)\nocite{Vaughan04,VF04,Gallo07}, while in others, variability remains nearly constant with energy (e.g., Ton S180 and NAB 0205+024; Vaughan et al. 2002 and Gallo et al. 2004, respectively)\nocite{Vaughan02,Gallo04}.  In the former cases, the change in variability strength is small, with a $<$ 10\% difference from 0.5 to 10 keV, so the bandpass effects at redshifts $z=0-1$ should remain small.  

We calculate the statistical error (i.e., measurement error) on the excess variance following Equation~11 of \citet{Vaughan03}.  
\begin{equation} err(\sigma^2_{\mathrm{nxs}}) = \sqrt{\left( \sqrt{\frac{2}{N}}\cdot\frac{\overline{\sigma^2_{err}}}{\bar{x}^2} \right)^2 + \left( \sqrt{\frac{\overline{\sigma^2_{err}}}{N}}\cdot\frac{2 \sigma_{\mathrm{nxs}}}{\bar{x}} \right)^2} \end{equation}
The large errors on $\sigma^2_{\mathrm{nxs}}$ (Table~2) are due to the small numbers of counts observed for most sources (e.g., 12 of 20 variable galaxies have $\lesssim$ 100 net counts); four variable galaxies with net counts $\lesssim$ 50 have excess variance measurements completely dominated by statistical uncertainty [$\sigma^2_{\mathrm{nxs}}$ $\lesssim$ err($\sigma^2_{\mathrm{nxs}}$)].  Nevertheless, most variable galaxies have excess variance measured at the $\gtrsim$1$\sigma$ level.  
%Although statistical error is a significant source of scatter on individual $\sigma^2_{\mathrm{nxs}}$ measurements, conclusions can be reached by considering the populations as a whole.  

The excess variance contains additional sources of uncertainty aside from statistical error: (1) random scatter intrinsic to the stochastic nature of AGN variability \citep{Vaughan03} and (2) uncertainty and systematic bias due to sparse sampling of the light curve.  The sparse pattern of \mbox{CDF-S} observations will lead to large uncertainty in the mean flux measurement, and since the measured mean will be closer to the sampled data points rather than the true mean, the variance measurements will tend to be underestimated \citep{Allevato11}.  We again employ a Monte Carlo simulation to model the uncertainty and systematic effects produced by random scatter and sparse sampling.  
%\footnote{Different realizations of the same stochastic process (i.e., different epochs of a continuous light curve) result in different measurements of the variance, even in the ideal case of a light curve with no measurement errors (see Vaughan et al. 2003 for further discussion).}

We follow the same procedure described above to produce 5,000 light curves for every variable source.  The mean and variance are calculated after sampling each simulated light curve with the \mbox{CDF-S} observing pattern.  The sampling bias can be corrected by rescaling the observed variance by a factor equal to the ratio between the ``true'' input variance (i.e., the observed variance used as input in the simulations) and the median output variance (i.e., the biased variance affected by sparse sampling): $f_{\mathrm{scale}}$ = $\sigma^2_{\mathrm{meas}}$/median($\sigma^2_{\mathrm{sim}}$). The median output variance is calculated over all 5,000 light curves.  The amount of systematic bias depends on the frequency and regularity of the sampling.  If the sampling is regular, the scaling factor will approach unity as the number of samples increases; however, the scaling factor will remain above unity even at high sampling frequency if the sampling is irregular \citep{Allevato11}.  The slope of the PSD will also affect the sampling bias --- a steeper PSD slope (i.e., $\beta = -2$ instead of $\beta = -1$) will result in a larger bias for a given sampling pattern. Since the intrinsic PSD slopes are not known, we apply uniform corrections assuming $\beta = -1$; however, source-to-source variations in PSD slope may be a significant source of scatter in variability measurements.

We find that the scaling factors range from $f_{\mathrm{scale}}$ $\approx$ 0.03 to 4.8, with a mean $f_{\mathrm{scale,mean}}$ $\approx$ 1.54 and a scatter on $f_{\mathrm{scale,mean}}$ of $\sigma_f\approx0.87$.  In $\approx$76\% of the sources, the scaling factor is greater than unity, indicating that the variance is underestimated due to the sampling bias.  Note that the Monte Carlo PSD is normalized by each source's variance, which is calculated using the source's light curve.  Since the heavy binning could smear out high frequency variations, this method could result in an artificially smaller scaling factor.  However, Allevato et al., in preparation\nocite{Allevato11}, find a similarly small bias ($f_{\mathrm{scale}}<2$) when sampling higher frequencies for a wide range of S/N ratios, gap lengths and sampling patterns.  More importantly, the scatter on the $f_{\mathrm{scale}}$ factor calculated for each source is large ($\approx 40\%-190\%$), so individual measurements, even when corrected for bias, are likely to be poor estimates of the intrinsic variance.  Therefore, variability properties of \mbox{CDF-S} galaxies are best considered in ensemble rather than on an individual basis.  Bias-corrected excess variances ($\sigma^2_{\mathrm{nxs,corr}}$) are listed in Table~2 and are used for all further analysis unless otherwise noted.
%scaling factor = median over 5,000 trials.  Scatter is calculated by taking the upper and lower quartiles.

%Applying $f_{\mathrm{scale}}$ to correct our measurements, we find that the sampling bias has no systematic effect on the luminosity-variability relation discussed in \S\ref{sec:sigma-Lx} (i.e., the slope and intercept measured before and after applying the correction factor are consistent within errors), although sparse sampling will introduce scatter into any correlation.  

%%%%%%%%%%%%%%%%%%%%%%%%%%%%%%%%%%%%%%%%%%%%%%%%%%%%%%%%%%%%%%%%%%%%%%%%%%%%%%%%%%%%%%%%%%%%%%%%%%%%%%%%%%%%
\subsection{Comparisons with XRB and ULX Variability}\label{sec:comparisons}

The three most likely sources of X-ray variability are \mbox{X-ray} binary (XRB) populations, ultraluminous \mbox{X-ray} sources (ULXs), and accreting SMBHs.  In this section, we show that the first two possibilities are not likely to dominate the measured galaxy variability.

To examine the potential contribution of \mbox{X-ray} binary populations to variability, we must first determine the relative contributions of low-mass \mbox{X-ray} binaries (LMXB) and high-mass \mbox{X-ray} binaries (HMXB) to the hard (2--10~keV), galaxy-wide \mbox{X-ray} luminosity ($L_{\mathrm{XRB}}$).\footnote{We measure variability in the 0.5--8~keV band, but by limiting the comparison to the 2--10~keV luminosities, we limit the contribution of other sources of galaxy-wide \mbox{X-ray} emission (hot gas, supernovae, supernova remnants, and O-stars), which fade sharply above 2~keV and can be considered negligible.}   Galaxy stellar mass ($M_\star$) scales the contribution of (older) LMXBs, and star formation rate (SFR) scales the contribution of (younger) HMXBs \citep[e.g., Equation 3 of][]{Lehmer10}.  $M_\star$ and SFR are calculated for each galaxy in \citet{Xue10} using the optical colors and the UV and IR luminosities.  Both the \citet{Lehmer10} relations and the \citet{Xue10} calculations adopt the same initial stellar mass function \citep{Kroupa01}.  However, since \citet{Lehmer10} and \citet{Xue10} use different formalisms for computing stellar masses (Bell et al. 2003 and Zibetti et al. 2009, respectively, which differ primarily due to their models of star formation history)\nocite{Bell03,Zibetti09}, we apply a correction factor of 2.6 to the stellar masses from \citet{Xue10}.  The SFR and $M_\star$ values for each variable galaxy are listed in Table~2.

By comparing the expected LMXB and HMXB contributions to the total luminosity, we can determine which population ought to dominate the variability.  In the variable galaxies, SFR ranges from 0.04 to 55 $M_\odot$ yr$^{-1}$ with a median of 2.2 $M_\odot$ yr$^{-1}$; $M_\star$ ranges from $2\times10^{7}$ to $6\times10^{11} M_\odot$ with a median of $1.2\times10^{10} M_\odot$. We find that all but six of 20 variable galaxies are expected to have a larger HMXB contribution.  $L_{\mathrm{HMXB}}$(SFR)/$L_{\mathrm{LMXB}}$($M_\star$) ranges from 0.01 to 75.8 with a median of 2.6 (Table~2).  HMXBs are generally more variable than LMXBs \citep{Gilfanov04}, so unless LMXBs dominate the \mbox{X-ray} output of a galaxy, we neglect their contribution.

To determine the variability expected from the HMXBs, we follow the relations in \S4.2.3 of \citet{Gilfanov04}, where the variability of the HMXB population is roughly determined by the galaxy's SFR.  The following relations were obtained from Monte Carlo simulations with a power-law HMXB luminosity function with slope $\alpha = -1.6$ and a cut-off luminosity at $L_{\mathrm{cut}}$ = 2$\times$10$^{40}$~ergs~s$^{-1}$:
\begin{equation} \frac{\sigma_{rms,tot}}{\sigma_{rms,0}} \sim 0.35^{+0.35}_{-0.10} \text{  for SFR $<$ 5 $M_{\odot}$ yr$^{-1}$} \end{equation}
\begin{equation} \frac{\sigma_{rms,tot}}{\sigma_{rms,0}} \sim 0.30^{+0.10}_{-0.05} \text{  for SFR = 5--10 $M_{\odot}$ yr$^{-1}$} \end{equation}
\begin{equation} \frac{\sigma_{rms,tot}}{\sigma_{rms,0}} = \frac{0.93}{\text{SFR}^{1/2}} \text{  for SFR $>$ 10 $M_{\odot}$ yr$^{-1}$} \end{equation}
Here, $\sigma_{rms,0}$ is the fractional rms (i.e., the square root of the excess variance) expected from an individual \mbox{X-ray} binary, which can be as large as 20--30\% on $\sim$year timescales \citep[e.g.,][]{Gilfanov10}, and $\sigma_{rms,tot}$ is the total variability.  We take $\sigma_{rms,0}=0.3$ and calculate the upper limit on $\sigma_{rms,tot}$.  In the most extreme case, we find that the upper limit on variability expected from an HMXB population is $\sigma^2_{\mathrm{XRB}} < 0.044$.

In six galaxies, LMXBs are expected to dominate the \mbox{X-ray} luminosity.  Four of the six have log $M_\star$ $>$ 10.5 $M_{\odot}$, and therefore have large enough stellar mass to follow a $\frac{\sigma_{rms,tot}}{\sigma_{rms,0}} \propto M_{\mathrm{stellar}}^{-1/2}$ law (\S4.4.2 of Gilfanov et al. 2004)\nocite{Gilfanov04}.  For the remaining two galaxies, we follow the trend in Figure 12 of \citet{Gilfanov04}.  The upper limit on variability expected from an LMXB population is $\sigma^2_{\mathrm{XRB}} < 0.02$.

\begin{figure}[t!]
\centering
\includegraphics[width=6.5in]{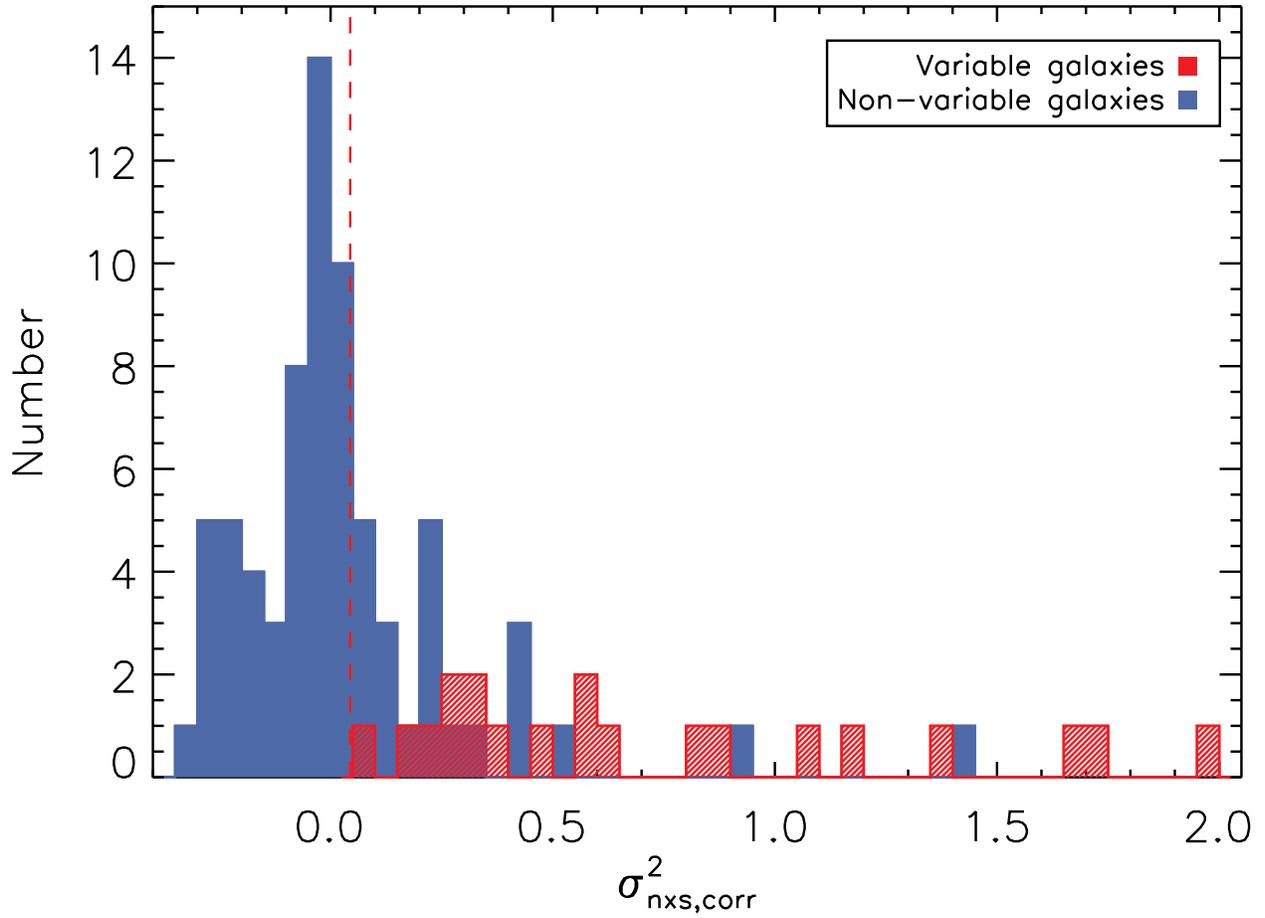}
\caption{The distribution of the bias-corrected normalized excess variance $\sigma_{\mathrm{nxs,corr}}^2$ for \mbox{CDF-S} sources classified as variable galaxies (red histogram), with non-variable galaxies (blue histogram) shown for reference.  As expected, the distribution of non-variable galaxies, though affected by scatter due to statistical fluctuations, centers around zero (i.e., variability strength is attributable to Poisson fluctuations). The red dashed line marks the maximum variability expected from an XRB population.}
\label{fig:factor_var}
\end{figure}

We find that XRB populations cannot explain the full extent of the \mbox{X-ray} variability for variable galaxies.  Figure~\ref{fig:factor_var} plots the distribution of $\sigma^2_{\mathrm{nxs,corr}}$ for \mbox{CDF-S} galaxies, showing that all the variable galaxies exhibit variability in excess of that expected from XRBs. The variable galaxies have a median $\sigma^2_{\mathrm{nxs,corr}}$/$\sigma^2_{\mathrm{XRB}}$ $\approx$ 42 (without the bias correction, $\sigma^2_{\mathrm{nxs}}$/$\sigma^2_{\mathrm{XRB}} \approx 14$), indicating that the contribution of an XRB population to the measured variability is small.  

%Note, that the comparison of individual sources with XRB variability remains valid, since the sampling bias is far more likely to result in an underestimate of the excess variance.
%Figure~\ref{fig:compare} compares the ratio of the measured normalized excess variance ($\sigma^2_{\mathrm{\mathrm{nxs}}}$) to the maximum variability that could be due to an XRB population ($\sigma^2_{\mathrm{XRB}}$) for variable (red histogram) and non-variable (blue histogram) galaxies.  

We also consider whether one or more ultraluminous \mbox{X-ray} sources (ULXs) may dominate a galaxy's \mbox{X-ray} output.  The nature of ULXs is still debated, but most likely involves accretion onto massive stellar black holes (30--100 $M_\odot$); a few cases may involve accretion onto intermediate mass black holes (100--300 $M_\odot$) or beamed emission from 10--20 $M_\odot$ black holes.  Typical luminosities span $L_{0.5-8 \mathrm{keV}}$ $\approx$ 10$^{39}$--10$^{41}$~ergs~s$^{-1}$ \citep[e.g.,][]{Lehmer06, Swartz11}.  In a survey of 1,441 \mbox{X-ray} point sources in 32 nearby galaxies in the \emph{Chandra} archive, \citet{Colbert04} found ULXs in 19 galaxies; the contribution of one or more ULXs to a galaxy's total \mbox{X-ray} point source luminosity ranged from 7\%--87\%, with a median contribution of 43\%.  
%(18 spirals, 9 ellipticals, and 5 irregular/mergers)   (10 spirals, 4 ellipticals, and 5 irregular/mergers)   %(56\%, 44\%, 100\%)

ULXs could potentially explain nine variable galaxies with $L_{0.5-8 \mathrm{keV}} < 10^{41}$~ergs~s$^{-1}$.  However, since ULXs tend to be associated with star-forming regions \citep[e.g.,][]{Swartz04} and occur more frequently in late-type/irregular galaxies than in early-type galaxies \citep[e.g.,][]{Walton11}, the five variable galaxies with both $L_{0.5-8 \mathrm{keV}} < 10^{41}$~ergs~s$^{-1}$ and late-type morphology (see \S\ref{sec:morph}) are more likely to host ULXs.  

\begin{figure*}[btp!]
\centering
\includegraphics[width=6in]{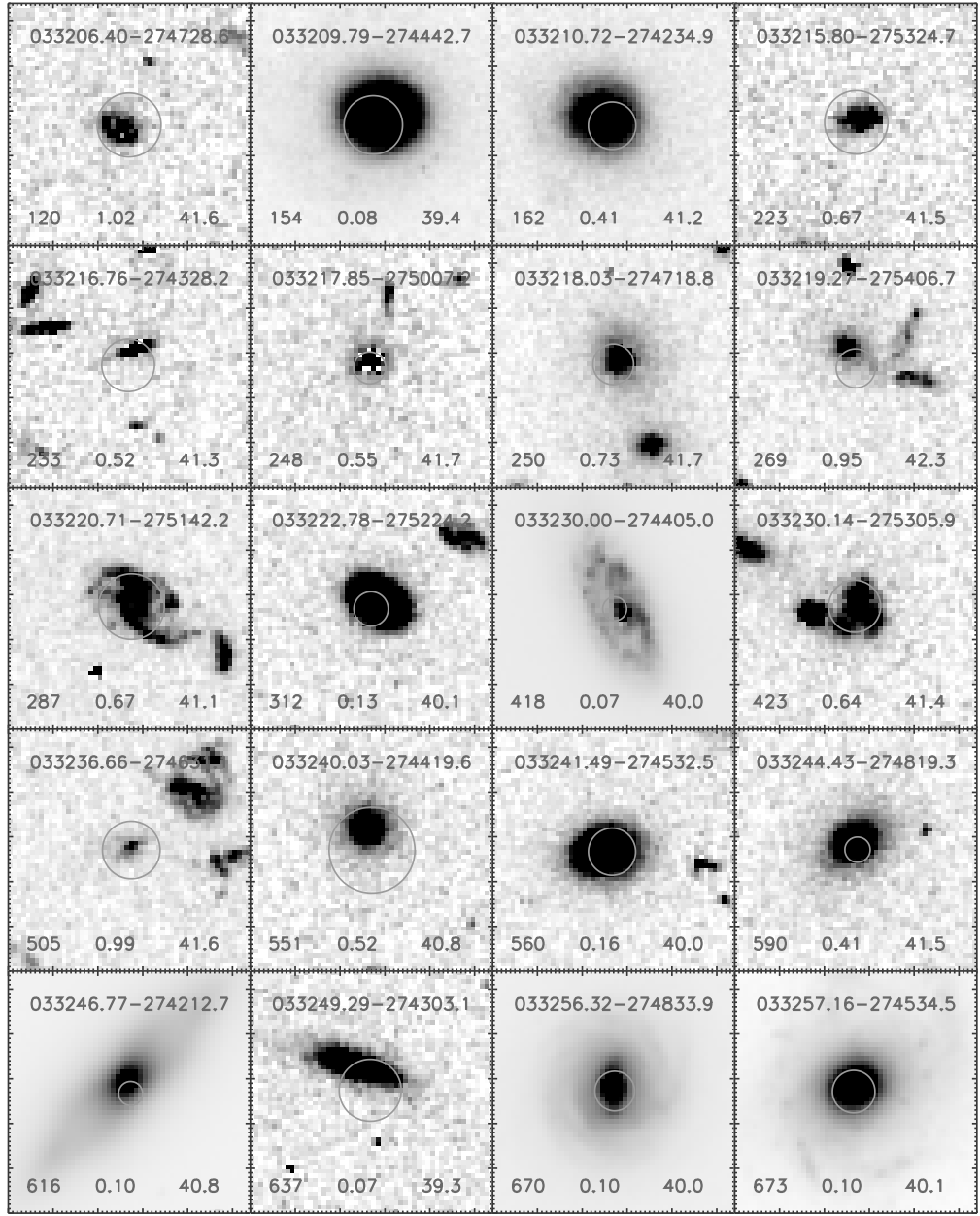}
\caption{Postage-stamp images from the GOODS-S/GEMS $HST$ $V606$ band for 20 variable galaxies. The label at the top of each image gives the source name. The numbers at the bottom of each image indicate the source number (XID) in the main \mbox{CDF-S} catalog, the adopted redshift, and the logarithm of the full-band luminosity as calculated in \S\ref{sec:overview}. The circle overplotted on each image has a radius of 1.5 times the \mbox{X-ray} positional error, which is calculated at the 90\% significance level, to illustrate whether an \mbox{X-ray} source is considered to be off-nuclear \citep{Lehmer06}.  Each image is 8$^{\prime\prime}$ $\times$ 8$^{\prime\prime}$ with the position of the \mbox{X-ray} source of interest located at the center.}
\label{fig:stamp_var}
\end{figure*}

One method of finding ULXs is to search for off-nuclear sources.  We plot postage-stamp images (8$^{\prime\prime}$ $\times$ 8$^{\prime\prime}$) of the GOODS-S/GEMS \emph{HST} $V606$-band for the variable galaxies in Figure~\ref{fig:stamp_var}.  The circle overplotted on each image has a radius 1.5 times the \emph{Chandra} positional erro, which is calculated at the 90\% significance level.  As in \citet{Lehmer06}, \mbox{X-ray} sources offset from the galaxy nucleus by more than 1.5 times the positional error are considered off-nuclear.  

We find one marginally off-nuclear source: 033219.27$-$275406.7 (XID = 269). The primary optical source appears to be an early-type galaxy with blending toward the galaxy to the lower right, suggesting a merger.  Both galaxies have similar redshifts ($z = 0.960$ and 0.956, respectively); the first redshift is spectroscopic and secure (see \S\ref{sec:overview}) and the second is photometric \citep{Xue10}.  The high \mbox{X-ray} luminosity ($L_{0.5-8 \mathrm{keV}}$ $\approx$ 2.2$\times$10$^{42}$~ergs~s$^{-1}$) suggests that a ULX is likely not the dominant source of \mbox{X-ray} emission from this galaxy.  

One variable galaxy, 033230.00$-$274405.0 (XID = 418), was previously identified as being off-nuclear in the 1 Ms \mbox{CDF-S} \citep{Lehmer06}.  With the additional data from the 4 Ms \mbox{CDF-S}, the X\nobreakdash-ray source position has been refined (with reduced uncertainty) to be consistent with the galaxy's nucleus.  The off-nuclear source discussed in the previous paragraph (XID = 269) was not detected in the 1 Ms \mbox{CDF-S}, so it was not considered by \citet{Lehmer06}.

We cannot rule out that a ULX may lie too close to a galaxy's nucleus to be detected as an off-nuclear source.  However, the possibility of ULXs in most variable galaxies is mitigated by high \mbox{X-ray} luminosities and/or early-type morphology, so since we find no plausible off-nuclear ULXs, we conclude that ULXs are unlikely to dominate the emission from variable galaxies.  Accretion onto a SMBH remains the best explanation of variable galaxies. 

%%%%%%%%%%%%%%%%%%%%%%%%%%%%%%%%%%%%%%%%%%%%%%%%%%%%%%%%%%%%%%%%%%%%%%%%%%%%%%%%%%%%%%%%%%%%%%%%%%%%%%%%%%%%
\section{Supporting Evidence for LLAGN}

We investigate the X-ray spectral shapes, the morphologies, and the optical spectral classifications of the variable galaxies for two purposes: (1) characterizing the variable galaxy population, and (2) determining whether their properties are consistent with those of LLAGN.  

%%%%%%%%%%%%%%%%%%%%%%%%%%%%%%%%%%%%%%%%%%%%%%%%%%%%%%%%%%%%%%%%%%%%%%%%%%%%%%%%%%%%%%%%%%%%%%%%%%%%%%%%%%%%
\subsection{\mbox{X-ray} Spectral Shape}\label{sec:stack}

As discussed in \S\ref{sec:overview}, \citet{Xue11} calculate the effective photon index for each source based on the ratio of count rates in the 2--8~keV and 0.5--2~keV bands (Table~2).  For 10 variable galaxies detected in the soft band but not the hard band, lower limits are listed.  For 4 low-count variable galaxies detected either in the full-band, soft-band, or both, no reliable effective photon index can be determined, so $\Gamma_{\mathrm{eff}}$ is set to 1.40.  

Since 14 sources have poorly determined $\Gamma_{\mathrm{eff}}$, we perform a stacking analysis on all variable galaxies with $<$~150 net counts to determine an average photon index for the sample.  The three highest count sources (with 199.2, 275.6, and 608.9 net counts) are excluded from the stacking analysis since they could dominate the stacked signal, but the results remain the same within errors if these sources are included.  

Following the procedure in \citet{Luo11}, we calculate the soft and hard-band counts in a 3$^{\prime\prime}$ diameter aperture for each source.  The background is calculated by averaging the counts in 1,000 randomly placed, source-free apertures within a 1$^\prime$-radius circle around the source position.  The individual source counts are summed and background is subtracted.  Aperture corrections are applied, averaged over all the observations weighted by exposure time (see Luo et al. 2011 for details)\nocite{Luo11}, before calculating the band ratio.  

The stacked effective photon index for the 17 relevant variable galaxies is $\Gamma_{\mathrm{stack}}$ $\approx$ 1.93$\pm$0.13, which is consistent with the typical photon index for local Seyfert galaxies ($\Gamma$ $\sim$ 1.8; e.g., Dadina et al. 2008)\nocite{Dadina08} at the 1$\sigma$ level.  Including all 20 variable galaxies, $\Gamma_{\mathrm{stack}}$ $\approx$ 1.82$\pm$0.08.  While absorption may still be present in some individual sources, the soft spectrum implied by the stacked effective photon index indicates that the average variable galaxy is not heavily obscured.  The X-ray luminosities are therefore intrinsically low and indicate that variable galaxies are most likely LLAGN.

%%%%%%%%%%%%%%%%%%%%%%%%%%%%%%%%%%%%%%%%%%%%%%%%%%%%%%%%%%%%%%%%%%%%%%%%%%%%%%%%%%%%%%%%%%%%%%%%%%%%%%%%%%%%
\subsection{Galaxy Morphologies}\label{sec:morph}

\begin{figure*}[btp!]
\begin{center}
\includegraphics[width=6.5in]{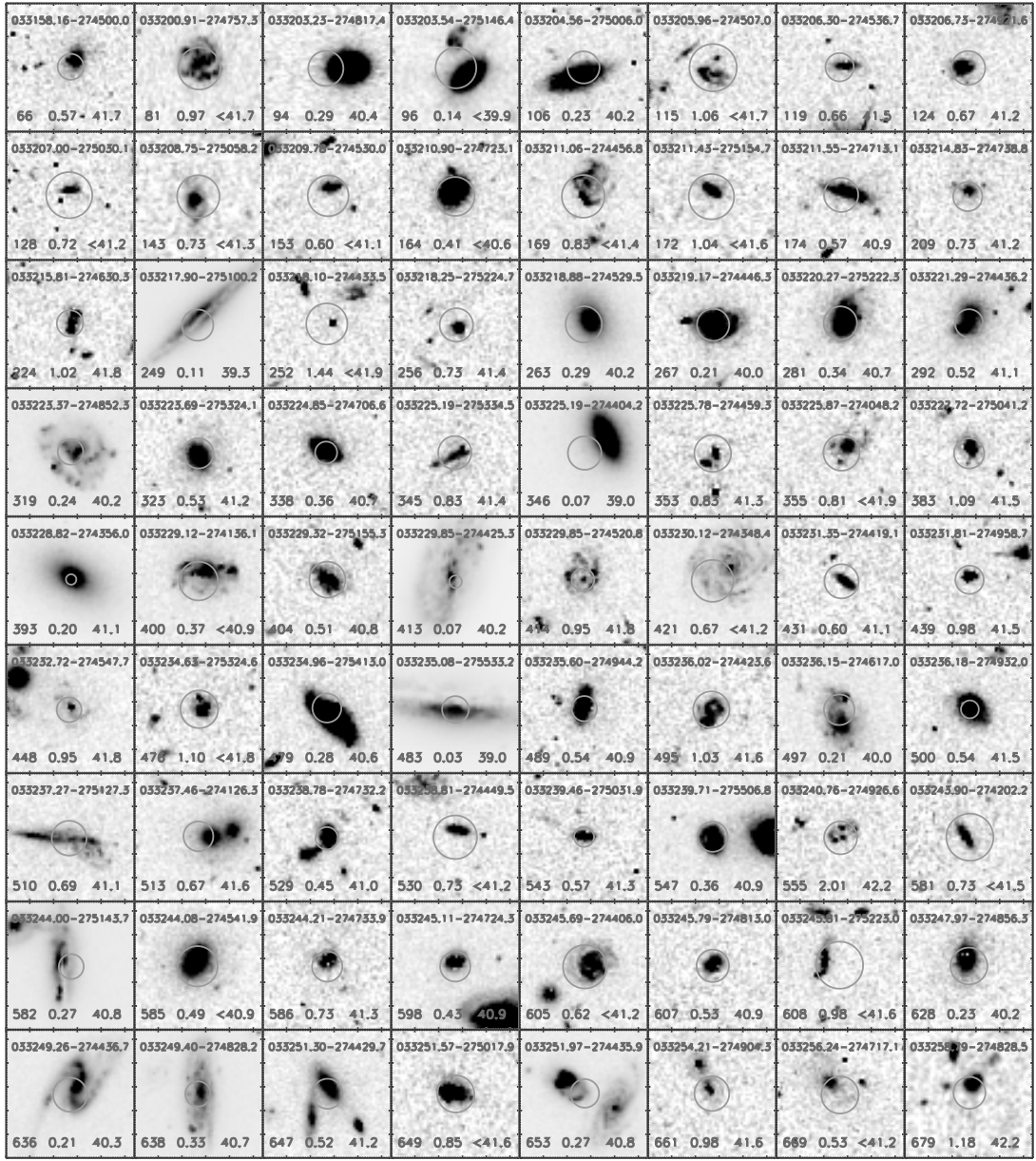}
\caption{Postage-stamp images from the GOODS-S/GEMS $HST$ $V606$ band for 72 non-variable galaxies.  The labels and error circles are the same as for Figure~\ref{fig:stamp_var}.}
\label{fig:stamp_non}
\end{center}
\end{figure*}

Postage-stamp images (8$^{\prime\prime}\times8^{\prime\prime}$) of the GOODS-S/GEMS \emph{HST} $V606$-band for \mbox{CDF-S} variable and non-variable galaxies are shown in Figures \ref{fig:stamp_var} and \ref{fig:stamp_non}, respectively, with \emph{Chandra} error circles overlaid.  Galaxies are classified by eye as late-type, early-type, irregular, or undetermined.  Mergers are also visually classified based on blending and/or tidal tails between two or more galaxies.  

Since visual classifications are subjective and are particularly unreliable for distant, poorly resolved galaxies, we also apply the color-magnitude relation given in \citet{Bell04}, where galaxies are considered part of the ``red sequence'' (i.e., early-type morphology) if they are redder than ($M_U$ -- $M_V$) = -0.31$z$ -- 0.08$M_V$ -- 0.51.  Galaxies with bluer colors are considered part of the ''blue cloud'' (i.e., late-type morphology).  Although the color-magnitude diagram leads to a more objective classification scheme, there are nevertheless significant uncertainties, both in the rest-frame magnitudes and colors, and in the definition of the \citet{Bell04} relation.  Moreover, the color classification is complicated by the fact that many of the galaxies lie in between the red sequence and blue cloud, in the so-called ``green valley.''  In cases where the color and the visually classified morphology disagree, we choose a final classification, preferring the visually classified morphologies in nearby, well-resolved galaxies, and the color classification in distant and/or poorly resolved galaxies. Considering only galaxies classified as early or late-type, the visual and color classifications agree $\approx$54\% of the time.  Table~2 lists each galaxy's morphological type as determined by visual classification, the rest-frame magnitude and color \citep[from][]{Xue11}, the color classification according to \citet{Bell04}, and the final classification.

Based on the final classification from Table 2, we find that variability does not prefer one morphology type over the other.  Variable galaxies have 40.0$^{+11.3}_{-9.5}$\% early-type morphology and 50.0$^{+10.6}_{-10.6}$\% late-type morphology, compared to 23.6$^{+5.7}_{-4.3}$\% and 51.4$^{+5.7}_{-5.8}$\%, respectively, in non-variable galaxies (the errors are 1$\sigma$ binomial errors).  If we instead apply only the \citet{Bell04} color classifications, then both the variable and non-variable galaxies prefer late-type morphologies (80.0$^{+6.0}_{-11.5}$\% and 72.2$^{+4.6}_{-5.8}$\%, respectively).  The $M_\star$ and SFR distributions for variable and non-variable galaxies show no significant difference ($P_{KS}$ $>$ 10\%).  

%A comparison between the SFR distributions for variable and non-variable galaxies shows a marginally significant difference (P$_{KS}$ $\approx$ 7.8\%).  The median SFR of variable galaxies is 2.2 $M_\odot$ yr$^{-1}$, compared to 5.2 $M_\odot$ yr$^{-1}$ for non-variable galaxies. The $M_\star$ distributions show no significant difference (P$_{KS}$ $\approx$ 16.2\%).
%Pierce et al. 2007 if more recent reference needed.

We find that the fraction of mergers among variable and non-variable galaxies is consistent within errors ($9.1^{+10.6}_{-3.1}$\% and $20.0^{+5.5}_{-3.9}$\%, respectively).  
%Mergers have been hypothesized to drive nuclear activity (e.g., Gunn 1979; Dahari 1984; Roos 1985; Taniguchi 1999), but recent studies of the \emph{Chandra} Deep Fields (Grogin et al. 2003; Grogin et al. 2005) show that asymmetry index distributions do not correlate with AGN activity.  The relatively fewer mergers in variable galaxies supports the view that mergers are not responsible for LLAGN activity.  On the other hand, mergers correlate with increased star formation activity (e.g., Schweizer 1982; Yun \& Hibbard 2001) and therefore are consistent with the presence of HMXB populations or ULXs, which may be responsible for the \mbox{X-ray} emission from the non-variable galaxies.  While increased star formation activity also correlates with AGN activity (e.g., Rafferty et al. 2011; Koss et al. 2011), these correlations are found for more luminous AGN ($L_{0.5-8 \mathrm{keV}}$ $>$ 10$^{42.5}$ ergs s$^{-1}$).  However, a fraction of the non-variable galaxy population may still host undetected AGN (e.g., due to small numbers of counts that prevent variability from being detected), which prevents firm conclusions on the merger-SF-AGN connection.  

%%%%%%%%%%%%%%%%%%%%%%%%%%%%%%%%%%%%%%%%%%%%%%%%%%%%%%%%%%%%%%%%%%%%%%%%%%%%%%%%%%%%%%%%%%%%%%%%%%%%%%%%%%%%
\subsection{Optical Spectroscopic Classification}

Most (18/20) variable galaxies have optical spectral observations \citep{Szokoly04, Zheng04, Mignoli05, Ravikumar07, Popesso09, Silverman10}, from which spectroscopic redshifts were determined.  In all cases, the optical spectra are classified as galaxies, showing only narrow emission lines or absorption lines.  

\citet{Szokoly04} classify objects in more detail.  Of the eight variable galaxies listed in the \citet{Szokoly04} catalog, two have typical galaxy spectra showing only absorption lines.  The remaining six are classified as having low-excitation emission lines consistent with H {\small II} region-type spectra. These objects would be classified as normal galaxies based on the optical data alone as the presence of the AGN cannot be established.  However, one of these, 033222.78$-$275224.2 (XID = 312), has sufficient signal-to-noise in the optical spectrum to measure line ratios.  This object is classified as a LINER by \citet{Szokoly04} via the line ratio diagnostics given by \citet{Ho93}. 

\section{Galaxy vs. AGN Variability}

%%%%%%%%%%%%%%%%%%%%%%%%%%%%%%%%%%%%%%%%%%%%%%%%%%%%%%%%%%%%%%%%%%%%%%%%%%%%%%%%%%%%%%%%%%%%%%%%%%%%%%%%%%%%
\subsection{The Variability-Luminosity Anti-Correlation}\label{sec:sigma_Lx}

\begin{figure}[btp!]
\centering
\includegraphics[width=5.5in]{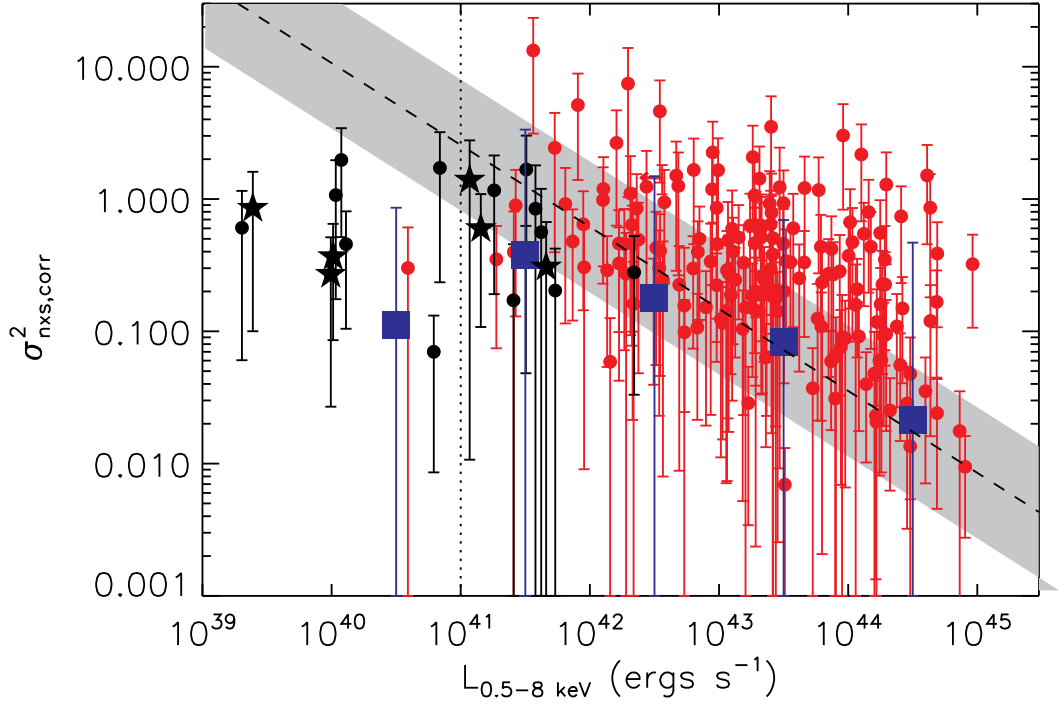}
\caption{The bias-corrected normalized excess variance ($\sigma^2_{\mathrm{nxs,corr}}$) vs. \mbox{X-ray} luminosity ($L_{0.5-8 \mathrm{keV}}$) for \mbox{CDF-S} sources classified as AGN (red circles) and variable galaxies (black circles and stars) in the 4 Ms main catalog, where sources with $L_{0.5-8 \mathrm{keV}}$$>$10$^{41}$~ergs~s$^{-1}$ are fitted with a weighted least-squares regression (dashed line); the shaded area shows the dispersion around the fitted line.  Black stars highlight six variable galaxies with significant XRB contribution to the total luminosity, discussed further in \S\ref{sec:suppressed}.  Error bars include both measurement errors and errors associated with red-noise scatter and sampling (\S\ref{sec:nxsvar}).  Large blue squares mark the weighted means for each luminosity bin; error bars are the standard deviation of the data in each bin. Ultraluminous \mbox{X-ray} sources (ULX) are generally not expected above $L_{0.5-8 \mathrm{keV}} \sim 10^{41}$~ergs~s$^{-1}$ (vertical dotted line).}
\label{fig:sigma_Lx}
\end{figure}

In this section, we investigate how galaxy variability compares to AGN variability.  We first confirm a significant anti-correlation between excess variance and \mbox{X-ray} luminosity among AGN, as seen in previous work \citep[e.g.,][]{BarrMushotzky86, LawrencePapadakis93, Nandra97, Hawkins00, Paolillo04}. The anti-correlation is plotted in Figure~\ref{fig:sigma_Lx} for significantly variable sources as $\sigma^2_{\mathrm{nxs,corr}}$ vs. $L_{0.5-8 \mathrm{keV}}$. The excess variance includes the bias correction discussed in \S\ref{sec:nxsvar}.  The rest-frame \mbox{X-ray} luminosities are calculated differently for AGN and galaxies, as described in \S\ref{sec:overview}. The Spearman rank correlation coefficient ($\rho_s$ = $-$0.44) shows the correlation is significant at $P_s$ $\approx$ 10$^{-9}$ (5.9$\sigma$) for AGN only (red circles).  The correlation increases in significance to $P_s \approx 8\times10^{-12}$ ($6.4\sigma$) if all variable sources above $L_{0.5-8 \mathrm{keV}}$ = 10$^{41}$~ergs~s$^{-1}$ are considered, including those classified as galaxies (black circles and stars). No significant correlation is found if the sample is limited to only the variable galaxies, which are discussed further in \S\ref{sec:suppressed}.  Note that the correlation coefficient and best-fit line parameters (given below) remain consistent within errors if calculated using the excess variance uncorrected for sampling bias ($\sigma^2_{\mathrm{nxs}}$). 

Since both variability and luminosity may depend on other parameters, such as black hole mass and/or accretion rate, we fit the data using {\sc sixlin.pro}, an IDL program adapted from \citet{Isobe90}.  A least-squares bisector fit to sources with $L_{0.5-8~\mathrm{keV}}$ $>$ 10$^{41}$~ergs~s$^{-1}$, weighted by the uncertainties in $\sigma_{\mathrm{nxs,corr}}^2$, results in $\sigma_{\mathrm{nxs,corr}}^2$ = (25.8$\pm$2.6)~$L_{0.5-8 \mathrm{keV}}^{-0.62\pm0.06}$. The slope and intercept are consistent within errors if the weights are not included.  Large blue squares in Figure \ref{fig:sigma_Lx} show the weighted means for each luminosity bin; as expected, these are consistent with the weighted least-squares bisector. Note that since weighted means will most closely follow data with small error bars, they are weighted heavily downward in this log-log plot. Limiting the sample further to sources with  $L_{0.5-8 \mathrm{keV}}$ $>$ 10$^{42}$~ergs~s$^{-1}$ results in $\sigma_{\mathrm{nxs,corr}}^2$ = (31.9$\pm$2.7)~$L_{0.5-8 \mathrm{keV}}^{-0.76\pm0.06}$.  The slopes are both consistent with the results of \citet{Nandra97}, where a weighted least-squares bisector fit results in $\sigma_{\mathrm{nxs}}^2$ $\propto$ $L_{2-10 \mathrm{keV}}^{-0.71\pm0.0.03}$ (Nandra et al. 1997\nocite{Nandra97} do not give the normalization of their relation).  This result is notable because the \citet{Nandra97} data sampled shorter (hour--day) timescales compared to the months--years timescales sampled by the \mbox{CDF-S}. 

The slope presented in this paper is significantly flatter than that found in Paolillo et al. (2004\nocite{Paolillo04}; $\sigma_{\mathrm{nxs}}^2$ $\propto$ $L_{0.5-8 \mathrm{keV}}^{-1.31\pm0.23}$), which included non-varying sources.  We choose not to include upper limits for non-varying sources in our analysis because the assumptions underlying survival analysis, which have been successfully applied to deal with censored data in other astronomical situations, do not apply here because: (1) the excess variance measurements of most sources lie near the detection limit, (2) a large percentage of sources do not have detected variability ($\approx$50\% of AGN and $\approx$78\% of galaxies), and (3) a significant percentage of sources with no detected variability, especially those classified as galaxies, may truly not be variable (i.e., $\sigma^2_{\mathrm{nxs}}$ $\sim$ 0).  By not including censored data in this paper's measurements, we likely bias the measured slopes and possibly the significance of the anti-correlation.  Nevertheless, the variability-luminosity anti-correlation has been observed to follow a model based on SMBH mass and accretion rate \citep[e.g.,][]{Papadakis08b}, suggesting that the anti-correlation is real, though probably not linear.  The model is discussed further in \S\ref{sec:model}.

We check for other potential biases that may affect the luminosity-variability anti-correlation.  The log-log plot of Figure~\ref{fig:sigma_Lx} has the disadvantage of ``hiding'' sources with negative $\sigma^2_{\mathrm{nxs,corr}}$ values. Sources with low flux values, and hence higher scatter in $\sigma^2_{\mathrm{nxs,corr}}$ will therefore appear to have stronger variability, since the $\sigma^2_{\mathrm{nxs,corr}}$ values scattered to negative values will be hidden. To check for this bias, we remove high scatter sources with $err$($\sigma^2_{\mathrm{nxs,corr}}$) $>$ 0.1, excluding all but 38 sources, and find that the $\sigma^2_{\mathrm{nxs,corr}}-$$L_{0.5-8 \mathrm{keV}}$ correlation remains significant at $P_s$ $\approx$ 0.9\% (2.6$\sigma$).  The best-fit slope and intercept remain consistent at the 1$\sigma$ level. 

The flux-limited nature of the \mbox{CDF-S} survey presents another potential bias.  Since luminosity correlates with redshift, and intrinsic variability timescales decrease with redshift, the decrease in variability strength as luminosity increases could, in principle, simply reflect the fact that shorter timescales are studied at higher redshifts, and therefore exhibit less variability due to the red-noise nature of AGN light curves. To check for a possible redshift bias, we examine the $\sigma^2_{\mathrm{nxs,corr}}$-$L_{0.5-8 \mathrm{keV}}$ correlation using a sub-sample within a narrow redshift range (0.55 $<$ $z$ $<$ 0.75).  This redshift range selects 10\% of the total sample, covers luminosities from $L_{0.5-8 \mathrm{keV}}$ $\sim$ 10$^{41.5}$ to $10^{44}$~ergs~s$^{-1}$, and results in minimal differences in rest-frame timescales.  The Spearman rank correlation coefficient for this sub-sample remains significant at $P_s$ $\approx$ 0.8\% (2.6$\sigma$); the slope and intercept are consistent at the 1$\sigma$ level. 

The above test also addresses another potential bias due to the redshift range of the sample.  The $\sigma^2_{nxs}$ values listed in Table 2 measure the variability strength in the observed frame, so they will sample different energy bands depending on the source's redshift (see \S\ref{sec:nxsvar} for discussion).  This could introduce bias if the variablity amplitude changes with increasing energy: from $0.5-8$ keV at $z$ = 0 to $1-16$ keV at $z$ = 1.  The slope and intercept remain the same within the narrow redshift range tested above, suggesting that any such bias does not have a significant effect on the sample.

%%%%%%%%%%%%%%%%%%%%%%%%%%%%%%%%%%%%%%%%%%%%%%%%%%%%%%%%%%%%%%%%%%%%%%%%%%%%%%%%%%%%%%%%%%%%%%%%%%%%%%%%%%%%
\subsection{Suppressed Variability in LLAGN}\label{sec:suppressed}

Variable sources with luminosities less than $L_{0.5-8 \mathrm{keV}}$ = 10$^{41}$~ergs~s$^{-1}$ tend to fall significantly below the extrapolated linear relation by factors of $\approx$6--80 (median factor of $\approx$24), indicating a drop in variability relative to the linear relation on long timescales for LLAGN.  This ``suppressed" variability can be shown to be intrinsic to properties of AGN variability rather than due to dilution by unrelated XRB populations.

In \S\ref{sec:comparisons}, we showed that galaxy variability cannot be attributed solely to XRB populations.  We now check whether the XRB contribution could nevertheless dilute the observed variability by estimating how much XRBs are expected to contribute to the total hard (2--10~keV), galaxy-wide \mbox{X-ray} luminosity ($L_{\mathrm{XRB}}$; see \S\ref{sec:comparisons}; Lehmer et al. 2010)\nocite{Lehmer10}.  

\begin{figure}[btp!]
\centering
\includegraphics[width=5.5in]{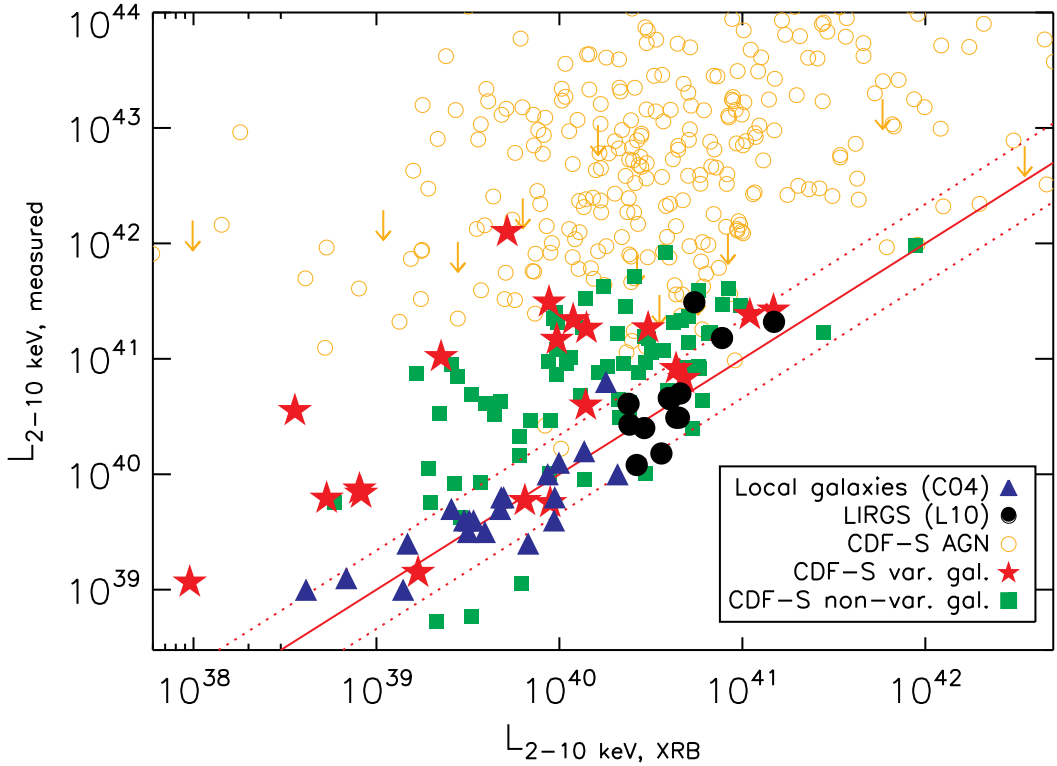}
\caption{The observed 2--10~keV \mbox{X-ray} power ($L_{2-10 \mathrm{keV, tot}}$) vs. that expected from an XRB population ($L_{2-10 \mathrm{keV, XRB}}$) for variable (red stars) and non-variable (green squares) galaxies.  Arrows designate upper limits on $L_{2-10 \mathrm{keV, tot}}$.  \mbox{CDF-S} AGN (open orange circles), 32 local galaxies \citep{Colbert04}, and 20 local luminous infrared galaxies \citep{Lehmer10} are plotted for reference.  Local galaxies lie on $L_{2-10 \mathrm{keV, tot}}$ = $L_{2-10 \mathrm{keV, XRB}}$ (red solid line), with a measured dispersion of 0.34 \citep[red dotted lines][]{Lehmer10}.}
\label{fig:compare_Lx}
\end{figure}

We compare $L_{\mathrm{XRB}}$ to the measured, intrinsic \mbox{X-ray} luminosity.  First, $L_{0.5-8 \mathrm{keV}}$ is converted to $L_{2-10 \mathrm{keV}}$, using the intrinsic photon index of $\Gamma$ = 1.8 adopted by \citet{Xue11} for AGN, and the stacked photon index $\Gamma_{\mathrm{stack}}$ $\approx$ 1.9 for galaxies (see \S\ref{sec:overview}). Figure~\ref{fig:compare_Lx} plots the measured 2--10~keV luminosity against that expected from an XRB population for variable galaxies (red stars) and non-variable galaxies (green squares).  For reference, we plot in the same figure 369 \mbox{CDF-S} AGN (open orange circles), 32 local galaxies \citep{Colbert04}, and 20 local luminous infrared galaxies, which are likely to be actively star-forming \citep{Lehmer10}.  The red solid line shows unity, while the dotted lines mark the dispersion observed in \citet{Lehmer10}.

A K-S test shows no significant difference between the relative XRB contribution in variable and non-variable galaxies.  The \mbox{CDF-S} galaxies on average tend to lie above unity, with a median $L_{2-10 \mathrm{keV}}$/$L_{\mathrm{XRB}}$ $\approx$ 4.6.  The \mbox{X-ray} excess in the non-variable galaxy population is perhaps not surprising given the flux-limited nature of the \mbox{CDF-S} survey, which is more likely to detect the objects at the bright end of the galaxy \mbox{X-ray} luminosity function, and may result in a high percentage of ``contamination" by AGN.  

%Another consideration is that since most galaxies are not detected in the hard band, we applied a uniform correction to the soft band flux in order to calculate the full-band luminosity (see \S\ref{sec:overview}) and this correction likely produces scatter on the $L_{2-10 \mathrm{keV}}$ measurement.

%For reference, we also calculate the ratio of total 2--10 keV luminosity to that expected from an XRB population for the Milky Way and Andromeda galaxies.  We take the SFR, $M_\star$, and $L_{2-10 \mathrm{keV, total}}$ from the literature (McKee \& Williams 1997 and Gilfanov et al. 2004 for the Milky Way; Barmby et al. 2008 and Supper et al. 1997 for M31).  The estimated ratios are 0.13 and 0.18, respectively, well below the $L_{2-10 \mathrm{keV, total}}$/$L_{2-10 \mathrm{keV, XRB}}$ = 3 cut-off.

The median \mbox{X-ray} excess for variable galaxies ($L_{2-10 \mathrm{keV, tot}}$/$L_{2-10 \mathrm{keV, XRB}}$ $\approx$ 9.2) suggests that XRBs contribute $\sim$11\% of the 2--10~keV luminosity for the average variable galaxy.  For six variable galaxies, the total X-ray emission is consistent with that expected from XRB emission within the scatter of the \citet{Lehmer10} relation (see Fig. \ref{fig:compare_Lx}), suggesting that dilution may be more significant in these sources.  Three of these sources have variability consistent with the linear variability-luminosity relation, while three have suppressed variability (filled black stars in Fig. \ref{fig:sigma_Lx}).  Dilution by XRB variability may therefore play a role in suppressed variability, but cannot fully explain the extent to which $\sigma^2_{\mathrm{nxs,corr}}$ is suppressed at low luminosities.  

%ULXs, which are also found to lie below the AGN-derived luminosity-variability relation (Gonzalez-Martin et al. 2011), are another potential source of dilution that could suppress variability at low \mbox{X-ray} luminosities.  However, as discussed in \S\ref{sec:comparisons}, there are no strong indications of ULXs in the variable galaxy population.

An alternative possible explanation for the suppressed variability at low luminosities is a change in accretion structure.  \citet{Ptak98} found a similar drop in variability strength below $L_{2-10 \mathrm{keV}}$ $\approx$ 2$\times$10$^{41}$~ergs~s$^{-1}$ in a sample of LLAGN and LINERs observed with \emph{ASCA} on variability timescales of less than a day. The authors hypothesized that a radiatively inefficient accretion flow (RIAF, e.g., Yuan \& Narayan 2004)\nocite{YuanNarayan04} could be responsible for suppressed short-timescale variability at low luminosities due to the larger extent of the \mbox{X-ray} source.  This scenario would not obviously explain the reduced variability on $\sim$month--year timescales seen here.  RIAF models also predict a hard X-ray photon index due to the lack of an optically thick accretion disk, which provides the soft X-ray photons.  The stacked X-ray photon index for variable galaxies ($\Gamma_{\mathrm{stack}}$ $\approx$ 1.93$\pm$0.13; \S\ref{sec:stack}) is inconsistent with this prediction.
%, although we cannot rule out that a change in accretion structure at low luminosities may nevertheless lead to lower variability if there is a change in the variability mechanism itself

Studies since \citet{Ptak98} have found evidence both for ``suppressed'' variability in LLAGN \citep[e.g.,][]{Ptak04,MarkowitzUttley05,Papadakis08a} and against it \citep[e.g.,][]{Binder09,Pian10}.  Similarly, objects such as narrow-line Seyfert 1 (NLS1) galaxies and the dwarf Seyfert NGC 4395 ($M_{BH}$ $\approx$ $3.6\times10^5 M_\odot$) exhibit ``excess'' vairability for their luminosity \citep{Boller96,Iwasawa10}.  However, when plotting variability against mass instead of luminosity \citep[e.g.,][]{Papadakis08a,Miniutti09}, such discrepancies disappear, with residual differences possibly due to varying accretion rates \citep[e.g.,][]{McHardy04,MarkowitzUttley05}.

To investigate the role of SMBH mass ($M_{BH}$) and accretion rate (normalized by the Eddington rate; $\dot{m}$ = $\dot{M}$/$\dot{M}_{Edd}$), we have obtained rough estimates for all variable sources.  Masses and Eddington ratios for all galaxies, variable and non-variable, are listed in Table 2.  The SMBH masses are estimated via the scaling relation between $M_{BH}$ and absolute $K$-band magnitude \citep{Graham07}: \begin{equation} \mathrm{log} \frac{M_{BH}}{M_\odot} = -0.37(\pm0.04)(M_K + 24) + 8.29(\pm0.08)\end{equation}
The total absolute rest-frame \mbox{$K$-band} magnitudes are derived from SED fitting in \citet{Xue10} with a random scatter of $\lesssim$0.3 mag.  An \mbox{X-ray} luminosity-dependent correction factor \citep[Equation~1 of][]{Vasudevan09} corrects for nuclear emission.  We assume that the host galaxy is bulge-dominated, a valid assumption for most AGN \citep[e.g.,][]{Kauffmann03, Grogin05}.  Several variable galaxies, however, are not bulge-dominated (nine are late-type; see Table 2), so their black hole masses may be overestimated.  We apply a luminosity-dependent bolometric correction ($\kappa_{2-10 \mathrm{keV}}$) to estimate the bolometric luminosity \citep{Marconi04} and calculate the Eddington ratio ($L_{bol}$/$L_{Edd}$ $\equiv$ [$\kappa_{2-10 \mathrm{keV}}$ $L_{2-10 \mathrm{keV}}$]/[$1.25\times10^{38}$~($M_{BH}$/$M_{\odot}$)]).  Note, that the \citet{Marconi04} correction is calculated explicitly for $L_{2-10 \mathrm{keV}}$ $>$ 10$^{42}$~ergs~s$^{-1}$; we extrapolate this relation down to the lower luminosities of the variable galaxy sample.

Both the $M_{BH}$ and $L_{bol}$ estimation techniques are known to have large dispersions.  The $M_{BH}$-$M_K$ scaling relation has a total scatter of 0.33 dex, and additional uncertainty will come from the luminosity-dependent correction for nuclear emission, which is based on template SEDs \citep{Vasudevan09}.  In addition, the assumption that all the variable galaxies are bulge-dominated will produce additional scatter.\footnote{The scaling relation in \citet{KormendyGebhardt01}, for example, has a much larger scatter of 0.56 dex largely because of poor bulge/disc separation \citep{Graham07}.}  The bolometric correction, too, has large scatter due to the intrinsic dispersion in the SED.  The uncertainty in $L_{bol}$ due to SED dispersion is $\sim$20\% for luminous AGN \citep{Elvis94,  Richards06}.  There is some debate regarding the similarity between the SEDs of LLAGN and luminous AGN (e.g., \S\ref{sec:intro}; Ho 1999 and Ho 2002 vs. Maoz et al. 2007)\nocite{Ho99,Ho02,Maoz07}; however, the luminous AGN SED dispersion likely serves as a lower limit to dispersion in LLAGN SEDs.

Variable galaxies tend to have lower accretion rates ($\langle\dot{m}\rangle$ = 4 $\times$ 10$^{-4}$) and masses ($\langle$$M_{BH}\rangle = 2.6\times10^7 M_\odot$) than variable AGN ($\langle\dot{m}\rangle$ = 9 $\times$ 10$^{-3}$; $\langle$$M_{BH}\rangle = 6.2\times10^7 M_\odot$), where we have limited the AGN sample to $z$~$<$~1 for purposes of comparison.  A K-S test shows that the differences in the $\dot{m}$ and $M_{BH}$ distributions are significant: P$_{KS}$ $\approx$ 1.7$\times$10$^{-5}$ and P$_{KS}$ $\approx$ 0.002, respectively.  The properties of variable AGN and galaxies are consistent with the range of estimates made by \citet{Babic07} for \mbox{X-ray} selected, $z$ $<$ 1 AGN in the 1 Ms \mbox{CDF-S}, which span $\dot{m}$~$\sim$~10$^{-5}-1$ (median $\approx$ 0.001) and $M_{BH} \sim 10^6~-~10^{10} M_\odot$ (median $\approx$ $10^8$ $M_\odot$).  

\begin{figure}[btp!]
\centering
\includegraphics[width=5.5in]{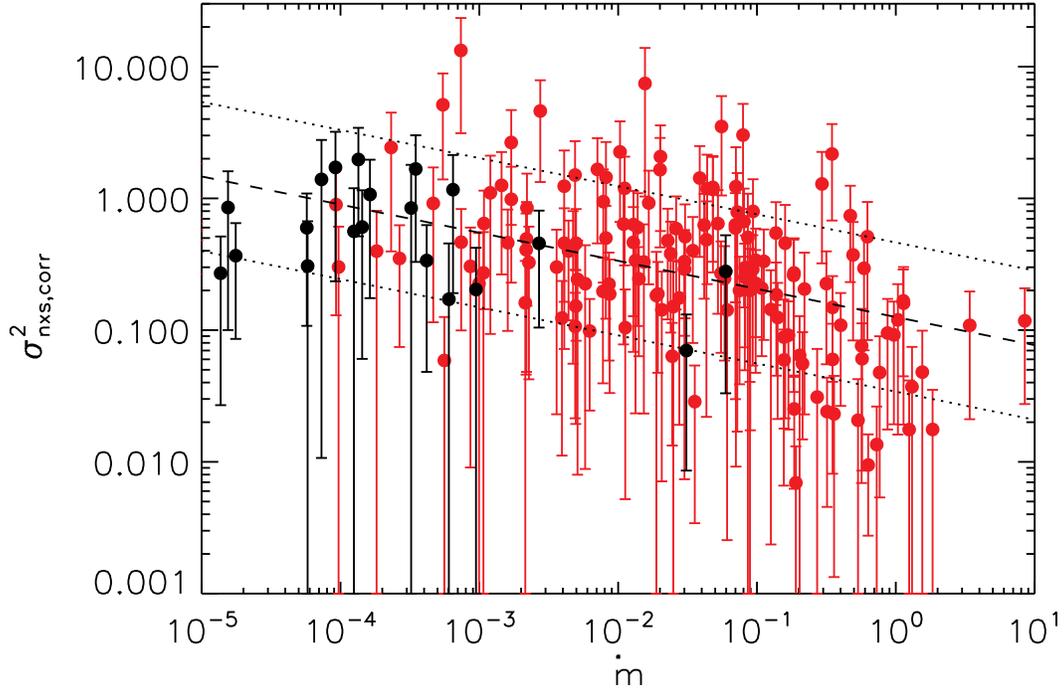}
\caption{The bias-corrected normalized excess variance ($\sigma^2_{\mathrm{nxs,corr}}$) vs. accretion rate normalized by the Eddington rate ($\dot{m}$= $\dot{M}$/$\dot{M}_{Edd}$) for \mbox{CDF-S} sources classified as AGN (red circles) and variable galaxies (black circles).  A weighted least-squares regression is fitted to all variable sources (dashed line), and the dotted lines show the dispersion around the best-fit line. }
\label{fig:mdot}
\end{figure}

Unlike previous studies \citep{O'Neill05,Papadakis08a,Miniutti09}, we find no significant (anti-)correlation between $\sigma^2_{nxs}$ and $M_{BH}$, most likely due to the large scatter in $\sigma^2_{nxs}$ and $M_{BH}$ measurements, combined with the narrow range of masses probed (since most sources lie between $M_{BH}$ $\sim$ 10$^7$--10$^9$ $M_\odot$).  It is not surprising, therefore, that we find a significant anti-correlation between $\sigma^2_{nxs}$ and $\dot{m}$ ($P_{KS} \approx 1.6\times10^{-11}$, $6.3\sigma$), plotted in Figure \ref{fig:mdot}, which is likely an artefact of the $\sigma^2_{nxs}$--$L_{0.5-8 \mathrm{keV}}$ anti-correlation.  Nevertheless, it is interesting that in Figure~\ref{fig:mdot} the variable galaxies (black circles) connect smoothly with the more luminous AGN (red circles), with no discrepancy in variability strength.  This suggests that the factor of 22.5 difference in $\dot{m}$ may explain the ``suppressed'' variability in variable galaxies.

%%%%%%%%%%%%%%%%%%%%%%%%%%%%%%%%%%%%%%%%%%%%%%%%%%%%%%%%%%%%%%%%%%%%%%%%%%%%%%%%%%%%%%%%%%%%%%%%%%%%%%%%%%%%
\subsection{Comparing the Variability-Luminosity Relation with Empirical Models}\label{sec:model}

A number of recent studies \citep[e.g.,][]{McHardy04,Papadakis04, O'Neill05, Papadakis08b} have shown that \mbox{X-ray} variability may be determined by a combination of $M_{BH}$ and $\dot{m}$, explaining the observed variability-luminosity relation.  AGN light curves appear to be universally described by a broken power-law PSD function, where the break frequency depends on mass and accretion rate: $\nu_{bf} = 0.029\eta\dot{m}$($M_{BH}$/$10^6 M_\odot$)$^{-1}$, where $\eta$ is the accretion efficiency, assumed to be $\eta = 0.1$ \citep{McHardy06}.  The excess variance is equivalent to the integral of the PSD between the minimum and maximum frequencies sampled by the data, so as long as the break frequency falls within this range, the excess variance at a given luminosity will decrease with increasing $M_{BH}$ and increase with increasing $\dot{m}$.

%We find a significant anti-correlation between $\sigma^2_{\mathrm{nxs,corr}}$ and $L_{bol}$/$L_{Edd}$ ($P_s$ = 4$\times$10$^{-11}$, 6.2$\sigma$ significance), but none between $\sigma^2_{\mathrm{nxs,corr}}$ and $M_{BH}$.  

We compare the $M_{BH}$ and $\dot{m}$ values estimated in \S\ref{sec:suppressed} to the variability measured for the variable AGN and galaxy populations.  Assuming a universal PSD function, we derive the variability-luminosity relations expected for the \mbox{CDF-S} sampling pattern, given a range of SMBH masses and accretion rates  \citep[e.g.,][and references therein]{Papadakis08b}.  The bolometric luminosity, which is calculated directly from $M_{BH}$ and $\dot{m}$, is converted to \mbox{X-ray} luminosity via a bolometric correction \citep{Marconi04}. Where the break frequency lies outside the timescales sampled by the data (depending on the combination of $\dot{m}$ and $M_{BH}$), the excess variance will remain constant.

\begin{figure}[btp!]
\centering
\includegraphics[width=5.5in]{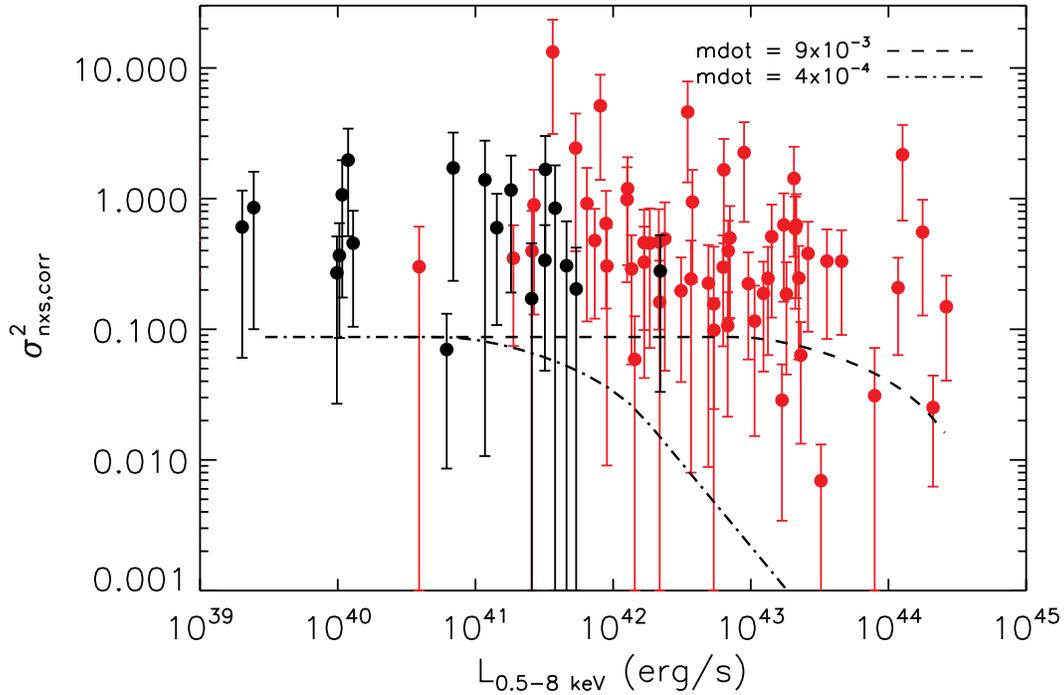}
\caption{The bias-corrected normalized excess variance ($\sigma^2_{\mathrm{nxs,corr}}$) vs. \mbox{X-ray} luminosity ($L_{0.5-8 \mathrm{keV}}$) as in Figure \ref{fig:sigma_Lx} for sources with $z$ $<$ 1.  Model relations are shown for a universal broken power-law PSD function, where the break frequency depends on $M_{BH}$ and $\dot{m}$.  The relations are calculated for $\dot{m}$ = 9 $\times$ 10$^{-3}$ (dashed line) and $\dot{m}$ = 4 $\times$ 10$^{-4}$ (dash-dotted line), where $M_{BH}$ = 10$^5$--10$^{10}$ $M_\odot$ (low to high $L_{0.5-8 \mathrm{keV}}$).}
\label{fig:sigma_Lx_physics}
\end{figure}

The model variability-luminosity relations are plotted in Figure~\ref{fig:sigma_Lx_physics} for the average accretion rates for variable ($\dot{m}$ = 9 $\times$ 10$^{-3}$; dashed line) and non-variable ($\dot{m}$ = 4 $\times$ 10$^{-4}$; dash-dotted line) galaxies covering $M_{BH}$ = 10$^5$--10$^10$ $M_\odot$ (low to high $L_{0.5-8 \mathrm{keV}}$).  Both variable populations are limited to $z$ $<$ 1 for comparison, and the relations are calculated at $z$ = 0.5 and $z$ = 0.7, the median redshifts for variable galaxies and AGN, respectively; using model redshifts $z$ = 0 or 1 resulted in negligible changes.  The shape of the observed variability-luminosity anti-correlation, including the plateau at low luminosity, is roughly reproduced by the model.  The plateau occurs when the break frequency lies outside the range of timescales sampled by the data.  Unfortunately, the long timescales (especially the long \emph{minimum} timescale of 4 Ms; see \S\ref{sec:nxsvar}) and the large scatter prevent the models from distinguishing between the significantly different accretion rates estimated for variable galaxies and AGN.  

Note that most sources are more variable than predicted by the model.  We note some possible sources of bias:  (1) The normalization of the PSD function used to calculate the models is based on a small sample of nearby AGN \citep{Papadakis04} and may therefore not be representative of \mbox{CDF-S} AGN out to $z \approx 1$.  (2) Both the models and the bias correction applied to the measured excess variance (\S\ref{sec:nxsvar}) depend on a universal broken power-law PSD function, but the slopes of the PSD may vary between individual sources, and some sources may even have a second break at shorter frequencies \citep[e.g.,][]{McHardy07}.  (3) Peculiar sources, such as NLS1s, exhibit higher variability\footnote{A notable exception is the NLS1 galaxy Akn 564, which is the only AGN observed to date to have a second, low-frequency break in its power spectrum, leading to lower variability on long timescales \citep{McHardy07}.}, so the variable sample may suffer from selection effects due to the higher sensitivity to more variable sources (\S\ref{sec:nxsvar}). 
\section{Conclusions and Future Work}

We have investigated the variability properties of sources classified as galaxies in the 4 Ms \mbox{CDF-S} by dividing the observations into $\approx$1 Ms epochs.  We find the following main results:

1.  Using a $\chi^2$-based Monte Carlo simulation appropriate for low-count sources, 20 of 92 galaxies are selected as variable, as well as 185 of 369 AGN.  Variability is effective in selecting AGN that do not meet other AGN selection criteria, such as cuts based on \mbox{X-ray} luminosity, \mbox{X-ray}-to-optical flux ratio, or excess \mbox{X-ray} emission compared to radio emission expected from star-forming galaxies (see \S\ref{sec:fluxvar}).  If all 20 variable galaxies are accreting SMBH, then the commonly used selection criteria employed by \citet{Xue11} miss 20 of 389 AGN ($\approx$5.1\%), and 19 of 54 LLAGN ($\approx$35.2\%; where LLAGN are arbitrarily defined as having $L_{0.5-8 \mathrm{keV}}$ $<$ 10$^{42}$ ergs s$^{-1}$).  Due to the low counts measured for most galaxies, it is likely that some non-variable galaxies may still host an AGN but not exhibit detectable variability, so these fractions serve as lower limits.

2.  We measure variability strength with the normalized excess variance, accounting for measurement error, red-noise scatter, and bias due to the \mbox{CDF-S} sampling pattern.  Comparing the bias-corrected excess variance to that expected from XRB populations, we find that XRBs cannot explain galaxy variability.  

3.  The possibility of energetically significant ULXs in variable galaxies is mitigated by high \mbox{X-ray} luminosities and/or early-type morphology.  In addition, we find no plausible off-nuclear ULXs.  

4.  Galaxy X-ray variability is most likely associated with accretion onto a relatively unobscured SMBH.  Though some absorption may be present, the individual and stacked photon indices show no indication of heavy obscuration and are consistent with the typical AGN spectral shape ($\Gamma_{\mathrm{stack}}$ $\sim$ 1.93$\pm$0.13).  
%SMBH accretion in variable galaxies is further supported by a preference toward early-type morphology.  
%Deeper follow-up spectroscopy would be useful to determine whether the other 19 variable galaxies have LINER characteristics.

5.  We confirm a significant anti-correlation between excess variance and \mbox{X-ray} luminosity, and find the slope and intercept to be consistent with what is reported in the literature for shorter timescales.  We show that the sampling bias induced by the \mbox{CDF-S} observing pattern does not affect the anti-correlation slope or significance.  

6.  Low-luminosity sources ($L_{0.5-8 \mathrm{keV}}$ $<$ 10$^{41}$~ergs~s$^{-1}$, largely variable galaxies) have ``suppressed'' variability compared to the extrapolated linear relation between excess variance and \mbox{X-ray} luminosity.  This may be explained by their lower accretion rates.  Variable galaxies sample a different mass-accretion rate space than the rest of the AGN population, with significantly lower accretion rates ($\langle\dot{m}\rangle$ = 4 $\times$ 10$^{-4}$) and masses ($\langle$$M_{BH}\rangle = 2.6\times10^7 M_\odot$) than variable AGN ($\langle\dot{m}\rangle$ = 9 $\times$ 10$^{-3}$; $\langle$$M_{BH}\rangle = 6.2\times10^7 M_\odot$) in the same redshift range ($z < 1$).  

7.  We find that an empirical model based on a universal broken power-law PSD function, where the break frequency depends on SMBH mass and accretion rate, roughly reproduces the shape of the variability-luminosity anti-correlation.  However, the normalization of the model is low compared to the data, suggesting systematic bias either within the data (e.g., selection effects) or the model (e.g., incorrect normalization for LLAGN at $z$ $\lesssim$ 1).

We have shown that the variability measured by deep X-ray surveys is an effective technique for selecting cosmologically distant LLAGN.  Extending the \mbox{CDF-S} to even longer exposures would enable detection of variability in both fainter and less variable sources, thus allowing better characterization of the properties and abundance of the LLAGN population.  Within the 4 Ms \mbox{CDF-S}, smaller time bins could be used to search for variability and better characterize the variability presented here.

In addition, follow-up optical spectroscopy is necessary to characterize the LLAGN population presented here.  While \citet{Szokoly04} were able to classify one object as a LINER, the spectra for the other sources did not have sufficient signal-to-noise to measure line ratios, so these objects would require either deeper optical spectroscopy or stacking of the currently available spectra.

%%%%%%%%%%%%%%%%%%%%%%%%%%%%%%%%%%%%%%%%%%%%%%%%%%%%%%%%%%%%%%%%%%%%%%%%%%%%%%%%%%%%%%%%%%%%%%%%%%%%%%%%%%%%
\acknowledgments

We thank P. Uttley, E. Feigelson, and C. Saez for helpful discussions on variability statistics.  We also thank M. Gilfanov for helpful discussions regarding XRB variability.  We thank the referee for constructive comments.  We acknowledge the financial support of NASA ADP grant NNX10AC99G (M.Y., W.N.B., Y.Q.X.) and Chandra X-ray Center grant SP1-12007A (W.N.B., Y.Q.X.).  M.P. acknowledges support from PRIN-2009 by the Italian MIUR.  F.E.B. acknowledges support from Programa de Financiamiento Basal
and CONICYT-Chile under grants FONDECYT 1101024, ALMA-CONICYT 31100004, and FONDAP-CATA 15010003.  B.D.L. acknowledges financial support provided by the Einstein Fellowship Program.  D.M.A. thanks the Science and Technology Facilities Council (STFC) for support.  

%%%%%%%%%%%%%%%%%%%%%%%%%%%%%%%%%%%%%%%%%%%%%%%%%%%%%%%%%%%%%%%%%%%%%%%%%%%%%%%%%%%%%%%%%%%%%%%%%%%%%%%%%%%%
%%%%%%%%%%%%%%%

%%%%%%%%%%%%%%%

%% Table %%
\begin{deluxetable}{ll}
\tabletypesize{\footnotesize}
\tablecolumns{2} 
\tablewidth{0pt} 
\tablecaption{Overview of Columns for Properties of \mbox{CDF-S} Galaxies}\label{tab:columns}
\tablehead{\colhead{Column}					&
	   \colhead{Description}				}
\startdata
1  & Sequence number in the \mbox{CDF-S} catalog \citep{Xue11} (i.e., XID) 			\\
2  & \mbox{CDF-S} name 										\\
3  & Redshift 											\\
4  & Method of redshift measurement (s = spectroscopic, p = photometric) 			\\
5  & Upper limit flag for net counts 								\\
6  & Total net counts in the 0.5--8.0~keV band or 3$\sigma$ upper limit 			\\
7  & log $L_{0.5-8 \mathrm{keV}}$ (ergs s$^{-1}$) 						\\
8-10  & Effective photon index ($\Gamma_{\mathrm{eff}}$)\tablenotemark{a} and corresponding errors 			\\
11  & Variability statistic ($X^2$) 								\\
12 & Probability that $X^2$ statistic is due to chance ($P_{X^2}$) 				\\
13 & Maximum-to-minimum flux ratio ($R_{\mathrm{max/min}}$) 					\\
14-15 & Excess variance ($\sigma_{\mathrm{nxs}}^2$)\tablenotemark{b} and corresponding error  				\\
16-17 & Bias-corrected excess variance ($\sigma_{\mathrm{nxs,corr}}^2$, see \S\ref{sec:nxsvar}) and corresponding error 	\\
18 & SFR ($M_\odot$ yr$^{-1}$) 									\\
19 & $M_\star$ (10$^{10}$ $M_\odot$) 								\\
20 & $L_{\mathrm{HMXB}}$/$L_{\mathrm{LMXB}}$ 							\\
21 & $\sigma_{\mathrm{XRB}}^2$ 									\\
22 & Visual galaxy classification 								\\
23 & $M_V$ 											\\
24 & $M_U - M_V$ 										\\
25 & Color galaxy classification 								\\
26 & Final galaxy classification 								\\
27 & log $M_{BH}$ ($M_{\odot}$) 								\\
28 & log $\dot{M}$/$\dot{M}_{Edd}$ 								\\
\enddata
\tablenotetext{a}{The effective photon index is calculated from the band ratio. For sources detected in the soft band but not the hard band, $\Gamma_{\mathrm{eff}}$ is a lower limit. When the counts are too low to determine reliably the photon index from the band ratio, $\Gamma_{\mathrm{eff}}$ is set to 1.4.}
\tablenotetext{b}{Errors on excess variance are calculated according to Equation (11) from \citet{Vaughan03}.  The excess variance is not significant for non-variable galaxies, but the errors can be used to calculate an upper limit.}
\end{deluxetable}

%% Table %%
\begin{deluxetable}{clrrrrrrrrr}
\tabletypesize{\footnotesize}
\tablecolumns{11} 
\tablewidth{0pt} 
\tablecaption{Properties of \mbox{CDF-S} Galaxies}\label{tab:Galvar}
\tablehead{\colhead{XID}							&
	   \colhead{CDF-S name}							&
	   \colhead{$z$\tablenotemark{a}}					&
	   \colhead{Net}							&
	   \colhead{log}							&
	   \colhead{$\Gamma_{\mathrm{eff}}$}					&
	   \colhead{$X^2$}							&
	   \colhead{$P_{X^2}$}							&
	   \colhead{$R_{\mathrm{max/min}}$}					&
	   \colhead{$\sigma_{\mathrm{nxs}}^2$}					&
	   \colhead{$\sigma_{\mathrm{nxs,corr}}^2$}				\\
	   \colhead{}								&
	   \colhead{}								&
	   \colhead{}								&
	   \colhead{counts}							&
	   \colhead{$L_{0.5-8 \mathrm{keV}}$}					&
	   \colhead{}								&
	   \colhead{}								&
	   \colhead{}								&
	   \colhead{}								&
	   \colhead{}								&
	   \colhead{}								}
\startdata
120 & 033206.40$-$274728.6  & 1.02  &     32.7   & 41.62  & 1.4	    		& 2.8  		& 0.0296 	& 4.1 		& 0.22$\pm$0.24	& 0.56$\pm$0.63 \\
154 & 033209.79$-$274442.7  & 0.08  &     46.9   & 39.39  & $>$0.9  		& 2.9  		& 0.0304 	& 4.1 		& 0.32$\pm$0.21	& 0.85$\pm$0.75\\
162 & 033210.72$-$274234.9  & 0.42  &     111.7  & 41.16  & 1.0$^{+0.4}_{-0.3}$	& 5.8  		& 0.0002 	& 3.4 		& 0.22$\pm$0.11	& 0.60$\pm$0.49\\
223 & 033215.80$-$275324.7  & 0.67  &  $<$43.7   & 41.51  & $>$1.0  		& 5.4  		& 0.0034 	& 8.1 		& 0.56$\pm$0.29	& 1.67$\pm$1.34\\
233 & 033216.76$-$274328.2  & 0.52  &     61.0   & 41.26  & $>$1.0		& 4.4  		& 0.0050	& 3.5 		& 0.43$\pm$0.21	& 1.16$\pm$0.97\\
\enddata
\tablecomments{Table 2 is presented in its entirety in the electronic edition. A portion is shown here for guidance regarding its form and content. The full table contains 28 columns of information for the 20 variable and 72 non-variable CDF-S galaxies.}
\tablenotetext{a}{For 18 of 20 variable galaxies and 61 of 72 non-variable galaxies, the redshift is measured spectroscopically and is ``secure'' (see \S\ref{sec:overview}). The remaining galaxies have photometric redshifts.}
\end{deluxetable}

%% Figure %%

%% Figure %%

%% Figure %%

%% Figure %%

%% Figure %%

%% Figure %%

%% Figure %%

%% Figure %%

%% Figure %%

%% Figure %%

%% Figure %%

%% Figure %%

%% Figure %%

%% Figure %%

\end{document}